\documentclass[traditabstract]{aa}
\usepackage{graphicx}
\usepackage{txfonts}
\usepackage{lipsum}
\usepackage{subcaption}         % necessary for continued figures, example in section 3                           % and appendix
\usepackage{lscape}             % to rotate a single page table, example in appendix.
\usepackage{placeins}           % useful with \FloatBarrier, to keep 
\usepackage{amsmath,siunitx}        
\usepackage[percent]{overpic} 

\usepackage{hyperref}
\begin{document}

   \title{Investigating Galactic Fountains in M101}
   \subtitle{Insights from Ionized, UV emission and Neutral Gas}

 %  \author{N. Puyaubreau\inst{1}
        % \and F. Déliat\inst{2}\fnmsep\thanks{Shows the usage of elements in the author field}
        % }

    \author{Aashiya Anitha Shaji\inst{1}
        \and
        Fran\c coise Combes\inst{1,2}   
        % \and 
        % Laurent Drissen\inst{3}
        \and
        Anne-Laure Melchior\inst{1}
        \and
        Anaëlle Hallé\inst{1}}

   \institute{Observatoire de Paris, LUX, CNRS, PSL University, Sorbonne University, 75014 Paris, France\\
             \email{aashiya.shaji@obspm.fr}
             %\thanks{Shows the usage of elements in the author field}
            \and Coll\`ege de France, 11 place Marcelin Berthelot, 75231 Paris, France\\ }

   \date{Received July 10, 2025}

% \abstract{}{}{}{}{}
% 5 {} token are mandatory
 
  \abstract
  {
  Spiral galaxy disks are thought to exist in a quasi-stationary state, between fresh gas accretion from cosmic filaments and disk star formation, self-regulated through supernovae feedback.
  % Aims
  Our goal here is to quantify these processes and probe their efficiency.
  % Methods
  While star formation can be traced at 10 Myr time-scales through H$\alpha$ emission, the signature of OB stars, and at 100 Myr scale with UV emission, the gas surface density is traced by \ion{H}{i} emission for the atomic phase.
  We choose to investigate feedback processes using fountain effects in M101, a nearby well-observed face-on galaxy. Face-on studies are very complementary to the more frequent edge-on observations of these fountains in the literature.  We use high-resolution data from THINGS for the \ion{H}{i} emission GALEX for UV, and SITELLE/SIGNALS IFU for the H$\alpha$ tracer.
  % Results
  We have identified 20 new \ion{H}{i} holes, in addition to the 52 holes found by Kamphuis in 1993. We study in more detail the nine holes satisfying strong criteria to be true fountain effects, compute their physical properties, and derive their energy balance. Only one small \ion{H}{i} hole still contains H$\alpha$ and young stars inside, while the largest hole of 2.4 kpc and oldest age (94 Myr) is deprived of H$\alpha$ and UV. 
  %Conclusions
  For face-on disks, the possibility to study simultaneously the \ion{H}{i} shell morphology, the stellar association, and kinematic evidence is of primordial importance. In M101, we have quantified how stellar feedback is responsible for carving the observed cavities in the atomic gas disk, and how it can expel above the disk the neutral gas, which is then unavailable for star formation during up to 100 Myr.}


   \maketitle

\begin{figure*}[ht!]
    \centering
    \begin{subfigure}[t]{0.49\textwidth}
        \centering
        \includegraphics[width=\textwidth]{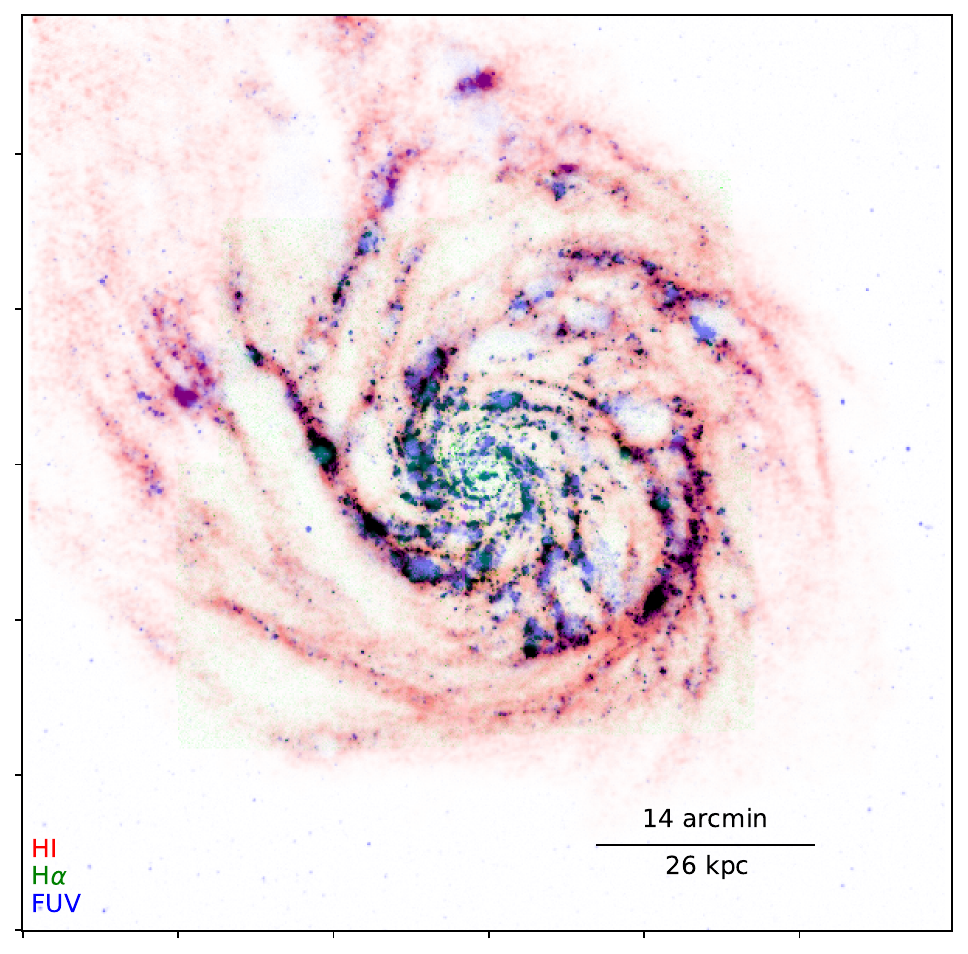}
        \caption{False-color RGB composite of M101 using \ion{H}{i} (THINGS, red), H$\alpha$ (SITELLE, green), and FUV (GALEX, blue). The color mapping is subtractive rather than additive, highlighting variations in neutral gas phase (H\textsc{i}), recently star-forming regions (H$\alpha$), and older star formation (FUV).}
        \label{fig:rgb_hi_ha_fuv}
    \end{subfigure}
    ~ 
    \begin{subfigure}[t]{0.49\textwidth}
        \centering
        \includegraphics[width=0.98\textwidth]{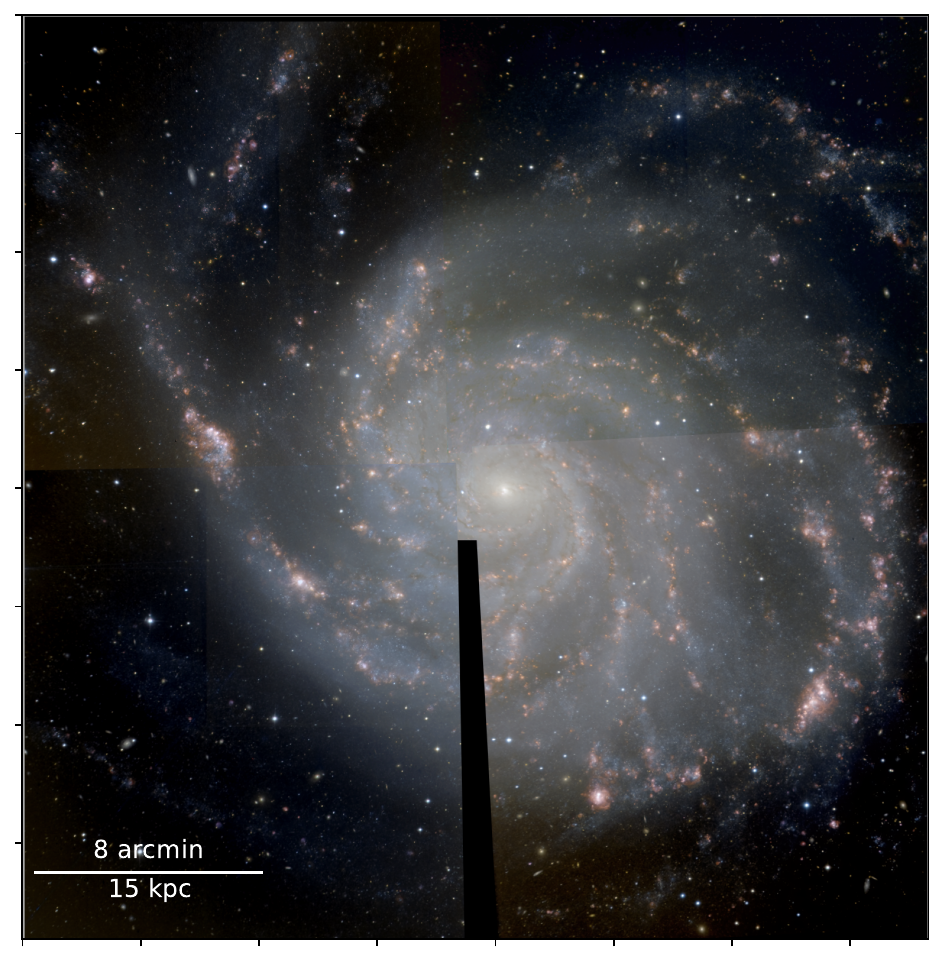}
        \caption{False-color RGB composite of M101 using SITELLE's SN1, SN2, and SN3 filters. The mosaic combines four pointings (each covering $11'\times11'$) to capture the full extent of the galaxy.}
        \label{fig:rgb_sitelle}
    \end{subfigure}%
    \caption{Multi-wavelength views of NGC 5457 or M101.}
    \label{fig:m101_rgb_comparison}
\end{figure*}
%%%%%%%%%%%%%%%%%%%%%%%%%%%%%%%%%%%%%%%%%%%%%%%%%%%%%%%%%%%%%%
\section{Introduction}
\label{sec:intro}

Feedback processes are pivotal regulators of galaxy evolution, governing the baryon cycle between stars, the interstellar medium (ISM), and the circumgalactic medium (CGM). The interplay of gas accretion, outflow, and recycling controls star formation rates, disk stability, and chemical enrichment \citep[e.g.,][]{2015Somerville}. Stellar feedback, primarily from supernovae (SNe), stellar winds, and intense radiation fields, can expel metal-rich gas from the disk. When this gas cools and condenses after ejection, it can fall back and fuel subsequent star formation, forming a closed-loop mechanism known as the galactic fountain \citep{1976Shapiro}.

In the galactic fountain framework, clustered supernova explosions and massive OB stellar winds heat the ISM to temperatures of order $\sim10^6$ K, launching gas vertically \citep[upto 1 kpc][]{1976Shapiro} at velocities that may or may not exceed the galaxy's escape speed. If the velocity is insufficient for escape, the ejected gas cools radiatively or mixes with cooler ISM phases, recombines into atomic hydrogen (\ion{H}{i}), and can later form molecular clouds (H$_2$) under favorable conditions \citep{1980Bregman}. These clouds, often seen as high-velocity complexes like those in the Milky Way \citep{1999VanWoerden} and M31 \citep{1975Davies,2007Westmeier}, eventually rain back onto the disk with velocities $\sim100$ km/s \citep{1976Shapiro} over timescales of $\sim$80-100 Myr \citep[33 Myr for cooling and another 47 Myr to return to the disk,][]{1990Houck}. This cycle contributes to disk turbulence, redistributes metals, and regulates star formation over galactic scales \citep{2016Armillotta}.

Galactic fountains have been observed most directly in edge-on galaxies such as NGC 891 \citep[$i \gtrsim 89^\circ$;][]{2007Osterloo} and NGC 2403 \citep[$i \sim 60^\circ$;][]{2002Fraternali}, where extraplanar \ion{H}{i} and ionized gas trace vertical fountain flows above the plane. However, nearly face-on systems like NGC 628 offer a complementary view. In such galaxies, vertical motions are largely hidden, but the in-plane manifestations—\ion{H}{i} holes, expanding ionized shells, and UV-luminous clusters—can be identified in the disk as the launch sites of fountain activity \citep{1992Kamphuis_Briggs}.

\begin{table}
    \centering
    \begin{tabular}{lcc}
    \hline
        Galaxy Properties & Values & References\\ 
        \hline
        Distance (Mpc)      & 6.4 & \citet{2011Shappee} \\
        Inclination (deg)   & 18 & \citet{2008Walter}\\
        Position Angle (deg)& 39 & \citet{Bottema1993}\\
        r$_{25}$ (arcmin) & 11.99 & \citet{2011Schruba}\\
        Scale length, $r^\mathrm{gas}_e$ (kpc) & 3.57 & \citet{2011Schruba}\\
        Scale length, $r^*_e$ (kpc) & 3.89 & \citet{2017Casasola}\\
        HI Mass (M$_\odot$) & 14.17 $\times 10^9$ & \citet{2008Walter}\\
        SFR (M$_\odot$ yr$^{-1}$) & 2.4 & \citet{2013Leroy}\\ 
        \hline
    \end{tabular}
    \caption{Properties of galaxy NGC 5457 or M101 from literature.}
    \label{tab:m101_props}
\end{table}
M101 (NGC 5457) is a nearby face-on spiral galaxy at a distance of $6.4\pm 0.7$ Mpc \citep{2011Shappee}, with a modest inclination of $18^\circ$ \citep{2008Walter}. It hosts vigorous star formation \citep[$\sim2.4~M_\odot$ yr$^{-1}$;][]{2013Leroy}, a flocculent barred structure, and no prominent bulge. At its distance 1 arcsec corresponds to 31pc, see Table \ref{tab:m101_props}. Early \ion{H}{i} studies by \citet{1993Kamphuis} on this galaxy, using $\sim$15 arcsec resolution Westerbork data, identified 52 \ion{H}{i} holes and two high-velocity complexes. One of these holes, located on the outer disk, was interpreted as a superbubble requiring $\sim1000$ SNe to account for its energy \citep{1991Kamphuis}. More recently, \citet{2011Chakraborti} discovered a distinct supershell in the inner disk of M101, traced by UV emission, likely powered by clustered supernovae. These studies established M101 as a promising case for investigating stellar feedback and the resulting ISM structures.

\section{Data Description}
\label{sect:data}
While \ion{H}{i}, UV, and ionized gas observations of M101 have each been studied independently, no systematic multiwavelength attempt has been made to identify galactic fountain activity and form a complete view of the feedback loop. This study presents the first such effort: using high-resolution data from THINGS (\ion{H}{i} 21 cm), GALEX (UV), and SITELLE/SIGNALS (optical IFU), we aim to identify observational signatures of galactic fountains in M101 from a face-on perspective.

The features we look for as signatures of galactic fountain activity span multiple wavelengths and gas phases. In the ultraviolet regime, we expect to detect compact star-forming complexes whose massive stars are capable of driving fountains via their stellar winds and supernova explosions. These appear as clumps in the GALEX FUV intensity maps, marking regions of recent massive star formation \citep{2004Keel}. While the time-scale for star formation is about 10 Myr for H$\alpha$ emission, it is of the order of 100 Myr for UV emission \citep[e.g.][]{Rampazzo2022}.

The supernovae from these associations inject energy into the surrounding interstellar medium, driving outflows of hot, ionized gas. Such ionized outflows can be traced through spectroscopic signatures (e.g., velocity-broadened H$\alpha$ emission lines) or as coherent morphological features such as shells and bubble-like structures in narrow-band emission maps \citep[][and references therein]{2018Rupke}.

These energetic events also displace the atomic hydrogen that previously occupied the region. As a result, observations in the 21 cm line of \ion{H}{i} often reveal these sites as gaps, depressions, or cavities in the atomic gas distribution. Such \ion{H}{i} holes are commonly seen in nearby spiral galaxies \citep{1986Brinks, 1990Deul, 1992Puche, 1993Kamphuis, 1998Wilcots,2002Fraternali, 2008Boomsma, 2011Bagetakos, 2020Pokhrel}, including the Milky Way \citep{1979Heiles, 1984Heiles, 1981Hu}. While some of these holes appear to be genuine voids, others are known to be filled with either hot ionized gas or colder molecular material. Their physical origin can vary, but many are linked to disk processes such as clustered supernova explosions, stellar winds, or in some cases, gas infall from the halo or intergalactic medium \citep[e.g.][]{Putman2012}.

Through this study, we aim to determine which disk cavities in M101 are consistent with feedback-driven fountains and to assess their sizes, possible ages, and implications for the cycling of gas between the disk and halo. We describe our data sources in Sect.~\ref{sect:data} and methodology in Sect.~\ref{sect:methods}. Our results are presented in Sect.~\ref{sec:results}, with their implications discussed in Sect.~\ref{sec:disc}.

% We adopt two complementary strategies. First, we revisit the \ion{H}{i} holes catalogued by \citet{1993Kamphuis} and evaluate them using a scoring system based on their \ion{H}{i} morphology and kinematics, UV emission, and ionized gas properties. Second, we identify shell- or bubble-like structures in SITELLE emission-line maps and trace them back to their potential UV and \ion{H}{i} counterparts. Through this dual approach, 

% %%%% maybe this can go in during the discussion and is not necessary at the introduction level?
% From the HI observations of 18 galaxies, \citet{2011Bagetakos} found that these fountains can create HI holes ranging from 100 pc to 2 kpc in size, with expansion velocities of 4-36 km/s and ages of 3-150 Myr. About 23\% of these holes were observed to lie outside $R_{25}$. Multiple fountain simulations indicate that the ejected material tends to fall back near its origin, maintaining the disk's radial chemical profile \citep{2009Melioli}.  \citet{1976Shapiro} states that the gas can travel upto a height of 1kpc and then fall back down with velocities of 100 km/s.  \citet{1990Houck} showed through hydrodynamic calculations that the cooling time for the hot gas is 33 million years and the time for newly formed clouds to return to the disk is 47 million years, neglecting drag forces.
% %%%% 

%%%%%%%%%%%%%%%%%%%%%%%%%%%%%%%%%%%%%%%%%%%%%%%%%%%%%%%%%%%%%%

This study makes use of multiwavelength datasets from GALEX, THINGS, and SITELLE to trace young stellar populations, atomic hydrogen structures, and ionized gas morphology and kinematics, respectively. Figure \ref{fig:rgb_hi_ha_fuv} is a hybrid map combining atomic gas, ionized gas, and ultraviolet emission.

\subsection{GALEX FUV and NUV Data}

The GALEX (Galaxy Evolution Explorer) archival data used in this study comes from the Guest Investigator Cycle 3 with ID: GI3\_050008\_NGC5457 in the FUV (1500–1800 Å) and NUV (1750–2800 Å) bands, with exposure lengths of 13293.4 and 13294.4 seconds, respectively. These data were retrieved from the MAST Archive and have spatial resolutions of 4.3 arcseconds (FUV) and 5.3 arcseconds (NUV). For this analysis, we use the intensity maps (fd-int.fits.gz and nd-int.fits.gz), which are provided in units of counts/sec/pixel with a pixel scale of 1.5 arcseconds. These images are used primarily for morphological identification of UV-bright stellar associations near HI holes and SITELLE shells.

\subsection{The HI Nearby Galaxy Survey}

The HI Nearby Galaxy Survey (THINGS), conducted using the NRAO Very Large Array (VLA), provides high-resolution 21 cm HI observations of nearby galaxies. We retrieve the data for M101 (NGC 5457) from the THINGS public archive\footnote{\url{https://www2.mpia-hd.mpg.de/THINGS/Data.html}}, along with its moment maps for integrated flux, velocity field, and velocity dispersion.

For our analysis, we used the data cube with natural (as it has low noise) weighting, which has a spectral resolution of 5.2 km s$^{-1}$ and a total of 69 velocity channels. We also use the maps obtained from the cube of robust weighting to cross-check hole boundaries and structure in ambiguous cases, as it has higher resolution but also higher noise levels. The image dimensions are 2048 $\times$ 2048 pixels, with a pixel size of $1.0''$. The RMS noise in one channel map is 0.46 mJy beam$^{-1}$, which corresponds to an HI column density sensitivity of $\sim 3.2 \times 10^{20}$ cm$^{-2}$. The synthesized beam size of the natural (respectively, robust) cube is $10.82'' \times 10.17''$ ($7.49''\times 6.07''$) with a position angle of $-67.0^\circ$ ( $-56.6^\circ$ measured north to east) \citep{2008Walter}. 

Table~\ref{tab:properties_comparison} compares the observational properties of the original \citet{1993Kamphuis} dataset with the THINGS data used in this study.

\begin{table}[h]
    \centering
    \begin{tabular}{l|c|c}
    \hline
    Properties                      & Kamphuis          & THINGS            \\  \hline
    Channel spacing (km s$^{-1}$)   & 4.1               & 5.2               \\  
    No. of Channels                 & 127               & 69                \\  
    Bandwidth (MHz)                 & 2.5               & 1.56              \\  
    Spatial Resolution            & 13''$\times$16''  & 7.49'' $\times$ 6.07'' \\  
    RMS Noise (mJy/beam)         & 0.5               & 0.46              \\  
    Exposure time                   & 16x12h            & 10h               \\  %(conf B+C+D)
    % Heliocentric velocity (km/s)    & 240               & 353, 106          \\  
    % NA: Bmaj, Bmin, BPA               & -                 & 10.82, 10.17, -67.0 \\ 
    %RO: 7.49 6.07 -56.6
    \hline
    \end{tabular}
    \caption{Comparison of HI observational setup or properties between Kamphuis and THINGS (NA). }
    \label{tab:properties_comparison}
\end{table}

\subsection{Star-formation, Ionized Gas, and Nebular Abundances Legacy Survey}

M101 is one of the galaxies observed in the Star-formation, Ionized Gas, and Nebular Abundances Legacy Survey (SIGNALS) \citep{2019Rousseau-Nepton}, conducted at the Canada-France-Hawaii Telescope (CFHT) using the SITELLE imaging spectrograph. Due to its proximity and extent, the galaxy was observed across four fields with three spectral filters: SN1 (363–386 nm), SN2 (482–513 nm), and SN3 (647–685 nm), at spectral resolutions of 1000 for SN1 and SN2, and 5000 for SN3. The data, retrieved from the Canadian Astronomy Data Centre (CADC), is available as HDF5 files for each filter and field.

The emission lines covered include [O II] $\lambda\lambda$3726, 3729 (SN1), H$\beta$ $\lambda$4861 and [O III] $\lambda\lambda$4959, 5007 (SN2), and H$\alpha$ $\lambda$6563, [N II] $\lambda\lambda$6548, 6584, and [S II] $\lambda\lambda$6717, 6731 (SN3). The deep frame, which comes with the datacube and is the sum of all individual interferometric frames, provides a high signal-to-noise image of the field per filter. A  mosaic using these narrowband-filters in the optical regime is shown in Fig.~\ref{fig:rgb_sitelle}. %This RGB image (Fig.~\ref{fig:rgb_sitelle}), which has been constructed with the deep frames of each filter, is used to identify bubble- or rim-like structures in our second approach.

We analyse the SITELLE data cubes using ORCS \citep{2015Martin}, a software designed for SITELLE data reduction and analysis. ORCS can be used to perform background subtraction and fit a single/multiple sincgauss profile to the entire cube, producing maps of velocity, flux, and velocity dispersion for each emission line. The instrumental line shape of SITELLE is a cardinal sine (sinc) function; hence, physical information is embedded in the resulting sincgauss (convolution of a sinc with a gaussian) profile \citep{2016Martin}.

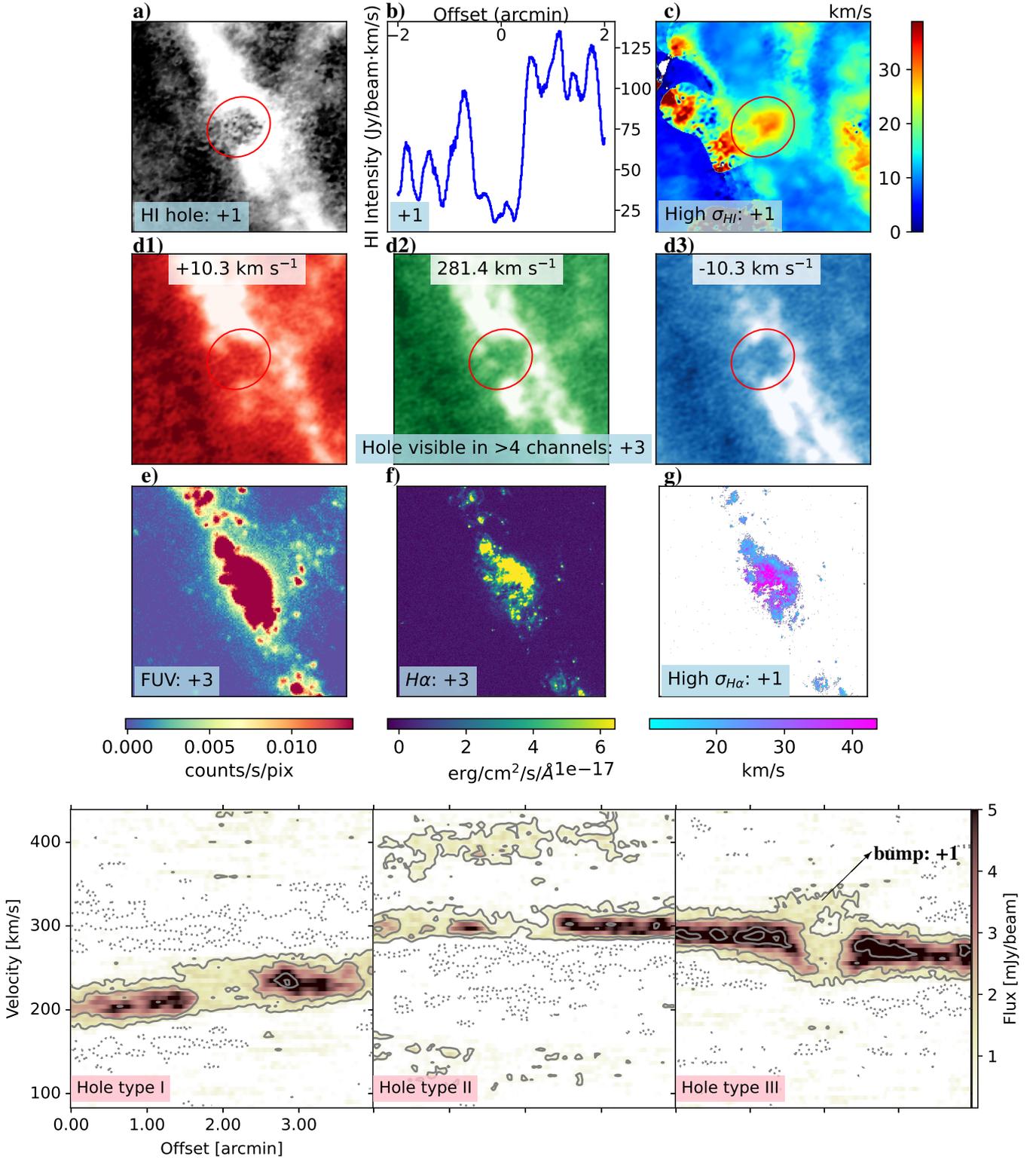
\begin{figure*}[!htb]
    \centering
    \begin{subfigure}[t]{\textwidth}
        \centering
        \begin{overpic}[width=0.8\linewidth,grid=false]{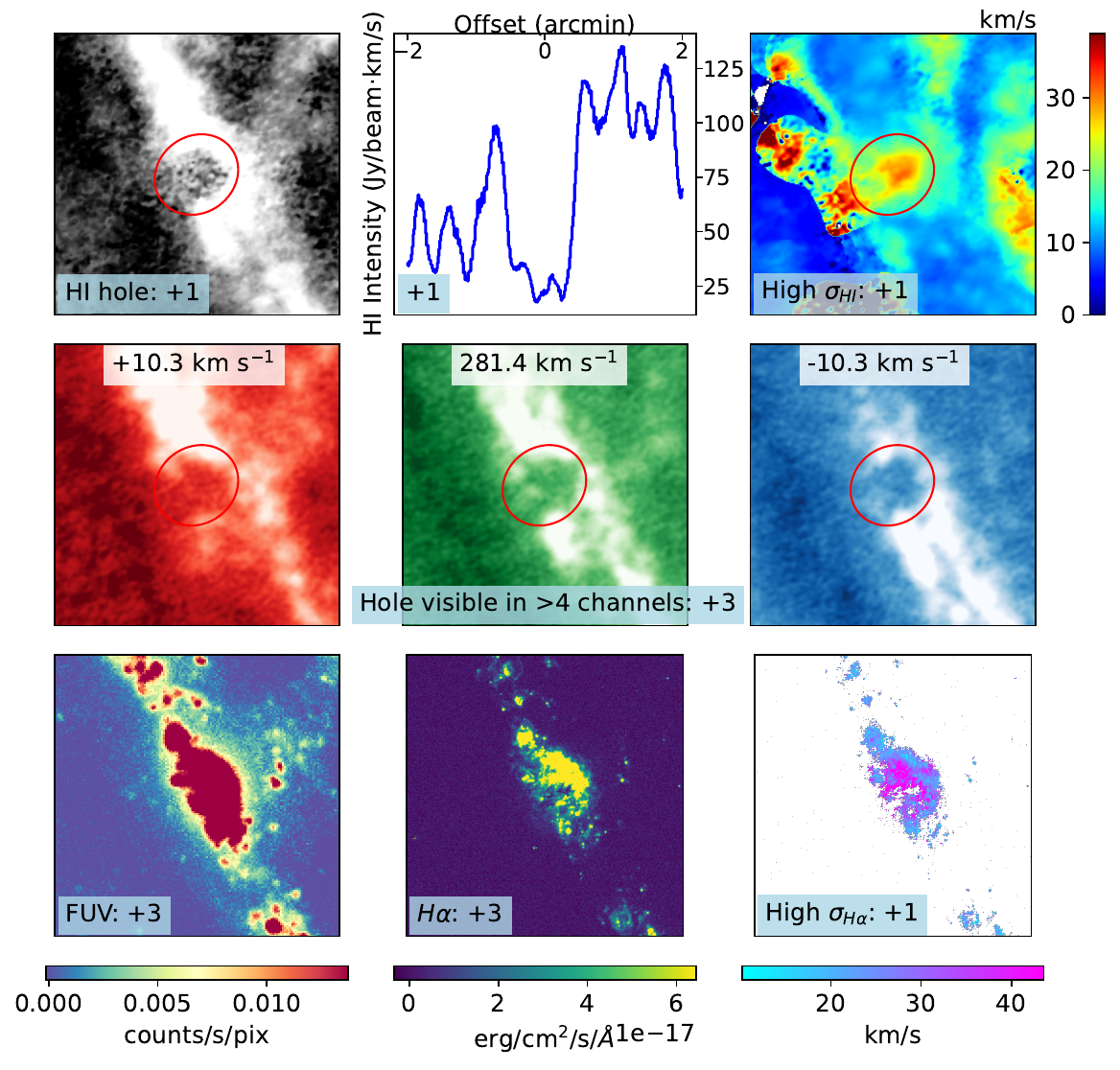}
            \put(5, 93){\large \bfseries a)}
            \put(35, 93){\large \bfseries b)}
            \put(68, 93){\large \bfseries c)}
            \put(5, 65){\large \bfseries d1)}
            \put(35, 65){\large \bfseries d2)}
            \put(68, 65){\large \bfseries d3)}
            \put(6, 37.5){\large \bfseries e)}
            \put(35, 37.5){\large \bfseries f)}
            \put(68, 37.5){\large \bfseries g)}
        \end{overpic}
    \end{subfigure}

    % Change subfigure label to 'h)'
    \renewcommand\thesubfigure{h}
    \begin{subfigure}[t]{\textwidth}
        \centering
        \begin{overpic}[width=0.98\textwidth,grid=false]{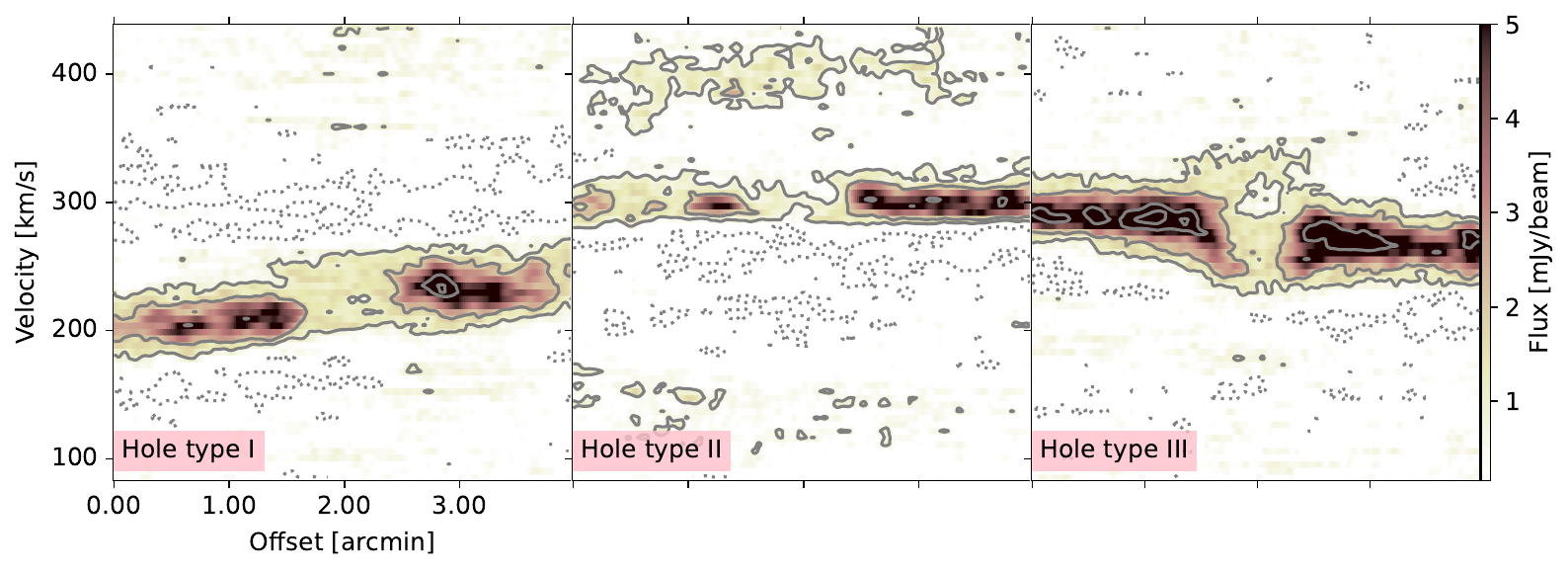}
            % Add arrow (adjust coordinates and size as needed)
            \put(80, 26){\vector(1,1){4.8}} % (x,y), direction, length
            \put(85, 30){\textbf{bump: +1}} % Annotation text
        \end{overpic}
        \caption{The three types of holes as per the position-velocity diagram, found in our sample. From left: Hole 3, Hole 2, Hole 1. Hole 2 suffers also from a tidal interaction, see Sec.~\ref{sec:results}.}
        \label{fig:PVtypes}
    \end{subfigure}
    \renewcommand\thesubfigure{\alph{subfigure})} % Reset to default numbering
    \caption{Pictorial depiction of the scoring criteria. The red ellipse in all the panels corresponds to Hole 1. The dashed blue line on the top left panel is the path across which the intensity profile and the rightmost position–velocity diagrams have been retrieved.}
    \label{fig:points}
\end{figure*}

% \begin{figure*}[!htb]
%   \centering
%   % wrap the includegraphics in an overpic environment
%   \begin{overpic}[width=0.8\linewidth,grid=false]{plots/Methodology_new.pdf}
%     % \put(xpct,ypct){LaTeX code}  coordinates run from (0,0) lower‐left to (100,100) upper‐right
%     \put(4, 98){\color{black}\large \bfseries a)}        % example text
%     \put(28, 98){\color{black}\large \bfseries b)}        % example text
%     \put(52, 98){\color{black}\large \bfseries c)}   % another label
%     
%     
%     \put(17, 26){\color{black}\large \bfseries h)}   % another label
%     % you can even put math
%     %\put(75, 25){\color{white}$I(\ell)$}                   
%   \end{overpic}
%   \caption{Pictorial depiction of the scoring criteria. The red ellipse in all the panels corresponds to Hole 1. The dashed blue line on the top left panel is the path across which the intensity profile and the position–velocity diagrams have been retrieved.}
%   \label{fig:points}
% \end{figure*}

\section{Methodology}
\label{sect:methods}

In this section, we describe how candidate regions hosting fountain-driven feedback features are identified using a multi-wavelength analysis of \ion{H}{i}, ultraviolet, and ionized gas data. Following \citet{2011Bagetakos} and \citet{2020Pokhrel}, we forgo automated \ion{H}{i}-hole detection in favor of visual inspection, owing to the irregular morphologies of \ion{H}{i} structures in galaxy disks. We identified candidate galactic fountain sites in M101 via two complementary routes. 

First, we revisit the 52 \ion{H}{i} holes catalogued by \citet{1993Kamphuis}, whose diameters range from $\sim0.8$ to $6.6$ kpc. These were originally detected using data with $\sim15$'' resolution, limiting sensitivity to structures larger than $\sim600$ pc. With the improved resolution of the THINGS dataset ($6''$, corresponding to $\sim200$ pc at M101’s distance of $6.4 \pm 0.7$ Mpc; \citealt{2011Shappee}), we revisit these holes and assess their physical nature using higher-quality \ion{H}{i}, SITELLE H$\alpha$, and GALEX UV data. To do so, we overlay the positions and extents of the Kamphuis holes as ellipses/circles on the THINGS \ion{H}{i} moment maps, SITELLE H$\alpha$ emission maps, and GALEX FUV intensity maps. Out of these 52 holes, one hole lies outside the THINGS field-of-view, so we drop it early on. The rest of these 51 structures are then subjected to a scoring scheme described in Sect.~\ref{ssect:scores}.

%In a nearly face-on disk like M101, stellar feedback signatures from FUV and H$\alpha$ should remain spatially coincident with the \ion{H}{i} hole if the structure is young or currently active. In contrast, inter-arm or fossil holes may lack such tracers.
In the second approach, we start from the SITELLE colour image given in Fig.~\ref{fig:rgb_sitelle}. We look for rim or bubble-like structures visually. We cross-match these with the FUV map to find young stellar clusters inside, usually lying exactly at their centres. Then we check the natural and robust integrated flux maps for the corresponding local minimum in the \ion{H}{i} maps. We also generated maps of [\ion{S}{ii}]/H$\alpha$, since shock-ionized shells (e.g. supernova remnants) often show elevated [\ion{S}{ii}]/H$\alpha > 0.5$. We identify 20 \ion{H}{i} holes in this manner. These holes are given points according to the scheme elaborated below. 

\subsection{Quality Assurance Scoring Scheme, Q}
\label{ssect:scores}

To assess the likelihood that each hole is feedback driven, we define a scoring scheme inspired by \citet{2011Bagetakos} and \citet{2020Pokhrel}. Each hole is assigned points based on its morphological and kinematic criteria as depicted in Figure.~\ref{fig:points}.

A +1 is assigned if the hole shows a well-defined flux minimum---quantified as a $\geq50\%$ drop in surface brightness---suggesting coherent depletion (Panels a and b of Fig.~\ref{fig:points}). Another +1 is added if this depression remains spatially consistent across at least three consecutive channels in the data cube, helping to mitigate false positives caused by turbulence or noise. Additional points are given depending on the number of contiguous velocity-channels where the depression is seen: 1–2 (5–10 km/s), 3–4 (11–20 km/s), and $>4$ ($>20$ km/s) channels receive +1, +2, and +3 points respectively. An extra point is added if the hole shows an elevated velocity dispersion in the \ion{H}{i} moment-2 map compared to the local surroundings (Panel c in Fig.~\ref{fig:points}). If the hole lies completely within a spiral arm, it receives +1, as spiral-arm holes are more likely to be star-formation-driven.

Next, we extract position–velocity (PV) diagrams along the spiral arm direction for each hole. In ambiguous cases, we rotate the cut across multiple angles (from $0^\circ$ to $180^\circ$) to optimize visibility. Based on the classification scheme of \citet{1986Brinks}, the holes are classified into three types as illustrated in Figure.~\ref{fig:PVtypes}. A type I is known as a fully blown out hole with neither receding nor approaching components visible on either side of the PV diagram (top panel). In type II holes, gas deviation is observable on one side the PV diagram (middle panel). Such cavities are called partially blown out holes. The third type is termed intact holes. Here, both receding and approaching sides are visible, forming an elliptical signature in PV space (bottom panel). Type III holes generally evolve into larger Type I holes over time. We assign +1 point for sharp PV edges in type I holes. Type II holes receive +1 if a well-defined receding or approaching ``bump" is visible. If this bump spans more than two channels, an additional +2 is given. This extended bump criterion, although rarely observed, was included for completeness.

To evaluate associated star formation, we assess UV (Panel e of Fig.~\ref{fig:points}) and H$\alpha$ (Panel f of Fig.~\ref{fig:points}) emission in and around each hole. If a UV-bright clump lies inside the \ion{H}{i} hole, we assign +2 points; if it lies near the edge, +1 point. We assign +2 if an H$\alpha$ clump lies inside, +1 if near the edge. Additionally, if the H$\alpha$ emission within the hole shows higher velocity dispersion (Panel g of Fig.~\ref{fig:points}) than its surroundings, we give it an extra point. Two extra points are given if the hole's edge contains bright rim or arcs (bubble-like structure) in the false-color RGB image created using SITELLE deep frames (Fig~\ref{fig:rgb_sitelle}), when scoring the holes found through approach one.

Each hole is thus scored across multiple observational dimensions: \ion{H}{i} morphology, kinematics, ionized gas features, and embedded stellar populations, out of a total of 20 points. Any structures with less than 10 points have been discarded, whereas those with more than 15 points are considered highly likely to host galactic fountains. We apply this score to all candidate holes; those scoring $\ge15$ are then characterized in detail through the following observed and derived parameters.

% \begin{figure*}
%     \centering
%     \includegraphics[width=0.9\linewidth]{plots/Methodlogy.pdf}
%     \caption{Pictorial depiction of the scoring criteria. The red ellipse in all the panels corresponds to Hole 1. The dashed blue line on the top left panel is the path across which the intensity profile and the position-velocity diagrams have been retrieved.}
%     \label{fig:points}
% \end{figure*}

\subsection{Estimation of Properties}

% \begin{figure}[!htb]
%     \centering
%     \includegraphics[width=0.95\linewidth, height= 14cm]{plots/pv_Types.pdf}
%     \caption{The three types of holes as per the position-velocity diagram, found in our sample. From top: Hole 3, Hole 2, Hole 1. Hole 2 suffers also from a tidal interaction, see Sec. \ref{sec:results}.}
%     \label{fig:PVtypes}
% \end{figure}
% \begin{figure*}
%     \centering
%     \includegraphics[width=\linewidth, height=3 cm]{plots/pv_Types_horizontal.pdf}
%     \caption{The three types of holes as per the position-velocity diagram, found in our sample. From left: Hole 2, Hole 3, Hole 1.}
%     \label{fig:PVtypes}
% \end{figure*}

\subsubsection{Observable Properties}
For each shell with points above 15, we record: central coordinates (RA, Dec; $\pm 3$"), heliocentric velocity $v_\textrm{Hel}$ (channel uncertainty $\pm5.2$ km/s), semi-major/minor axes $b_\textrm{maj}$, $b_\textrm{min}$, axial ratio $b_\textrm{min}/b_\textrm{maj}$, position angle (±30°), Brinks type, expansion velocity $v_\textrm{exp}$ (see below), and mean \ion{H}{i} flux density ($S_\textrm{avg}$). 

We take $v_\textrm{Hel}$ to be the velocity channel at which the hole is clearly visible (Panel d2 of Fig.~\ref{fig:points}). The calculation of expansion velocity depends on the type of hole. For type I holes, $v_\textrm{exp}$ cannot be determined, but its upper limit can be deduced from the average velocity dispersion within the hole. For partially blown-out holes, $v_\textrm{exp} = v_\textrm{hel} - v_\textrm{bump}$ where $v_\textrm{bump}$ is the velocity at the bump (where the gas is deviating from the surroundings, rightmost figure in Fig.~\ref{fig:PVtypes}). For the intact holes, we use the average difference between $v_\textrm{hel}$ and the velocities of the gas on the approaching and receding sides of the hole. The uncertainty of the calculation is again the velocity resolution (5.2 km/s). We determine the $S_\textrm{avg}$ by averaging the flux values on the integrated \ion{H}{i} map over a region of size twice the $b_\textrm{maj}$ and $b_\textrm{min}$ of that of the hole itself. The errors here are estimated to be of the order 10\% arising from the uncertainty about the position of the hole center.

\subsubsection{Derived Properties}
\label{sssect:derived}
The derived properties are as follows. Most of the following equations, if not specifically cited, are adopted from the past \ion{H}{i} hole studies conducted by \citet{2011Bagetakos} and \citet{2020Pokhrel}:
\begin{enumerate}
    \item Diameter, $d = 2\sqrt{b_\textrm{maj} b_\textrm{min}}$
    
    \item The galactocentric distance in pc of a hole with centre ($\alpha, \delta$) to the galaxy centre ($\alpha_0, \delta_0$) in radians, can be obtained using the distance to the galaxy ($D$) in pc, position angle ($\theta$), and inclination ($i$):
    \begin{align}
        R &= D \sqrt{(x'')^2 + (y'')^2} \\
        \textrm{where } x'' &= x\sin\theta + y\cos\theta \notag \\
        y'' &= \frac{y\cos\theta - x\cos\theta}{\cos i} \notag \\
        \textrm{and } x &= (\alpha - \alpha_0)\cos\delta_0 \notag \\
        y &= \delta - \delta_0 \notag
    \end{align}
    
    \item The kinetic age in Myr can be calculated from diameter in parsec and expansion velocity in km/s,
    \begin{equation}
        t_\textrm{kin} = 0.978 \frac{d/2}{v_\textrm{exp}}
    \end{equation}
    \item \ion{H}{i} column density, 
    \begin{equation}
        N_\textrm{HI} = 1.823 \times 10^{18} \sum_i \left[ \frac{S^i_\textrm{avg}}{1.66\times 10^{-3} B_\textrm{maj} B_\textrm{min}} .\Delta v \right]
    \end{equation}
    Here, $S^i_\textrm{avg}$ is the mean \ion{H}{i} flux density around the hole in mJy/beam, in each velocity channel $i$. $B_\textrm{maj}, B_\textrm{min}$ are beam sizes in arcsec. $\Delta v$ channel width in km/s.
    \item The effective thickness, $\ell(r)$, of the atomic hydrogen disk is calculated from the inclination of the galaxy, $i$, as:
    \begin{equation}
        \ell(r) [\mathrm{pc}] = \frac{Z_0 \sqrt{2\pi}}{\cos i}
    \end{equation}
    Here, the scale height $Z_0$ is derived from the \ion{H}{i} velocity dispersion ($\approx 12$ km/s), assuming hydrostatic equilibrium 
    of the atomic gas in the gravitational potential of the stellar disk, with negligible dark matter \citep{1997Combes}:
    \begin{equation} \label{eqn:z0}
        Z_0 [\mathrm{pc}] = \frac{v_\mathrm{disp}}{\sqrt{4\pi G \rho(r)}}
    \end{equation}
    where
    \begin{align}
        \rho(r) = \frac{\Sigma^*(r)}{h_*(r)} = \frac{\Sigma_0^* e^{-r/r_e^*}} {h_*}
    \end{align}
   is the volumic stellar density, $\Sigma^*(r)$ the stellar surface density, $r^*_e$ the stellar exponential radial scale and $h^*_z$ is the constant scale height of stellar disks \citep[e.g.][]{Bottema1993}. We adopt the stellar exponential radial scale from \citet{2017Casasola}, which utilised the IRAC data to calculate the stellar scale length of M101 to be 2.09 arcmin $\approx 3.89$ kpc. The $\Sigma_0^*$ can be deduced to be 381 M$_\odot$/pc$^2$ from their study. We consider $h_* = r_e/7.3 \approx 0.53$ kpc following \citet{2008Leroy}, who adopted the average flattening ratio of 7.3 from \citet{2002Kregel}. This is based on the assumption that the height of the stellar disk is independent of its radius \citep{1982VanderKruit}.

    \item The energy required to power the hole can be calculated using two different formulae. First using \citet{1974Chevalier}, which is more dependent on the diameter of the hole, and the second using  \citet{1987McCray}, which has high dependence on the expansion velocity. 
    \begin{align} 
        E_{\mathrm{Ch}} [\mathrm{erg}] &= 5.3 \times 10^{43} n_0^{1.12}[\mathrm{cm}^{-3}] \left ( \frac{d [\mathrm{pc}]}{2} \right)^{3.12} v_{\mathrm{exp}}^{1.4} [\mathrm{km s}^{-1}] \label{eqn:Ech} \\
        E_{\mathrm{Mc}}[\mathrm{erg}] &= n_0 \left( \frac{d[\mathrm{pc}]}{194}\right)^2 \left( \frac{v_\mathrm{exp}}{5.7} \right)^3 \times 10^{51} 
    \end{align}

    Here we take the density of the ambient medium, $n_0$, as the particle density of the atomic hydrogen by ignoring contributions from He and H$_2$. The midplane \ion{H}{i} volume density ($n_\textrm{HI}$) is calculated using the column density as follows:

    \begin{equation}
        n_\textrm{HI} = \frac{N_\textrm{HI}}{3.08\times10^{18} \ell(r)}
    \end{equation}
    
    \item \ion{H}{i} Mass of the hole: 
    \begin{equation}
        M_\mathrm{HI} [\mathrm{M}_\odot] = 0.0245 n_\mathrm{HI} V
    \end{equation}
    where the volume of the hole is given by
    \begin{equation}
        V [\textrm{pc}^3] = \frac{4}{3}\pi \left ( \frac{d}{2} \right)^3
    \end{equation}
\end{enumerate}
% \begin{figure}
%     \centering
%     \includegraphics[width=0.95\linewidth]{plots/PV_Paths.pdf}
%     \caption{Position-velocity (PV) diagram paths overlaid on the moment-zero map of HI data. The paths, aligned parallel to the minor axis, are labelled with their respective offsets from the center near their starting positions. Each path has a width of 5 pixels. The 51 HI regions identified by \citet{1993Kamphuis} are shown in white.}
%     \label{fig:HI_PV_paths}
% \end{figure}

% \begin{figure*}
%     \centering
%     \includegraphics[ width=\linewidth]{plots/PV_plots.pdf}
%     \caption{Position-velocity diagrams corresponding to the paths shown in the previous figure. Distance offsets from the galaxy centre are labelled in the top left of each subplot. The colour scale in the position-velocity diagrams is set between the 10th and 98th percentiles of the data distribution to enhance contrast. Black crosses indicate the RA positions of HI holes; their y-positions hold no physical significance and are placed at the average value of each subplot for visualization purposes.}
%     \label{fig:HI_pv}
% \end{figure*}

% \begin{figure*}
%     \centering
%     \includegraphics[ width=\linewidth]{plots/PV_plots_95.pdf}
%     \caption{Same as \ref{fig:HI_pv} with the colour scale in the position-velocity diagrams is set between the 10th and 95th percentiles of the data distribution to enhance faint features. }
%     \label{fig:HI_pv_95}
% \end{figure*}
    \begin{table*}
    \centering
    \begin{tabular}{lcccccccccc}
    \hline
    Name & RA & Dec & $b_{maj}$ & $b_{min}$ & PA  & $b_{min}/b_{min}$ & $S_\mathrm{avg}\Delta v$ & Type & $v_\mathrm{exp} $& $v_\mathrm{hel}$ \\
    & hh:mm:ss & dd:mm:ss & arcsec & arcsec & degrees & & Jy/beam * m/s & & km s$^{-1}$ & km s$^{-1}$\\
    \hline

    Hole 1 & 14:03:54.84 & +54:21:41.78 & 34 & 30 & 37 & 0.9 & 86.7 & 1 & 54 & 275 \\
    Hole 2 & 14:03:41.69 & +54:29:38.29 & 33 & 25 & 65 & 0.8 & 54.9 & 2 & 41 & 296 \\
    Hole 3 & 14:03:27.13 & +54:17:35.74 & 45 & 33 & 0  & 0.7 & 86.2 & 3 & 13 & 224 \\
    Hole 4 & 14:02:25.19 & +54:27:44.45 & 30 & 23 & 84 & 0.8 & 132.0 & 3 & 28 & 255 \\
    Hole 5 & 14:03:33.88 & +54:18:29.59 & 9 & 8 & 90 & 0.9 & 91.3 & 0 & 9 & 248 \\
    Hole 6 & 14:02:47.53 & +54:14:54.31 & 17 & 6 & 90 & 0.4 & 132.4 & 0 & 10 & 183 \\
    Hole 7 & 14:02:17.63 & +54:18:31.60 & 13 & 8 & 90  & 0.6 & 167.8 & 0 & 8 & 183 \\
    Hole 8 & 14:02:37.50 & +54:28:57.82 & 20 & 7 & 259 & 0.4 & 96.1 & 0 & 7 & 265 \\
    Hole 9 & 14:02:14.72 & +54:25:47.95 & 12 & 6 & 90  & 0.5 & 57.2 & 0 & 9 & 235 \\
    \hline
    \end{tabular}
    \caption{The physical properties observed in the nine \ion{H}{i} holes that are sites hosting galactic fountains. Brinks type of the smaller holes have been set to zero considering its position-velocity diagrams are too ambiguous.} 
    \label{tab:GF_physical}%

  \vspace{-0.0cm} %space between tables

  \caption{The derived properties of the nine \ion{H}{i} holes that are sites hosting galactic fountains.}
    \begin{tabular}{lcccccccc}
    \hline
    Name & Diameter & Galactocentric  & Kinetic   & $N(HI)$ & Effective   & E$_\mathrm{mc}$ & E$_\mathrm{ch}$ & Mass \\ 
         &          &  Distance       &  Age      &         &  thickness  &                  &              & Displaced \\
    & kpc & kpc & Myr & $10^{20}$ cm$^{-2}$ & pc & erg & erg & M$_\odot$ \\    \hline
    
    Hole 1 & 2.03 & 12.76 & 18.8 & 8.6 & 31.1 & 8.37e+56 & 3.96e+56 & 9.71e+08 \\
    Hole 2 & 1.80 & 19.45 & 21.9 & 5.5 & 13.2 & 4.31e+56 & 2.91e+56 & 1.01e+09 \\
    Hole 3 & 2.43 & 8.33 & 93.6 & 8.6 & 55.0 & 9.47e+54 & 5.00e+55 & 9.44e+08 \\
    Hole 4 & 1.67 & 20.43 & 29.9 & 13.2 & 11.6 & 3.24e+56 & 4.19e+56 & 2.22e+09 \\
    Hole 5 & 0.55 & 8.35 & 30.6 & 9.1 & 54.9 & 1.71e+53 & 3.10e+53 & 1.16e+07 \\
    Hole 6 & 0.68 & 14.19 & 33.8 & 13.2 & 25.9 & 1.09e+54 & 2.40e+54 & 6.62e+07 \\
    Hole 7 & 0.69 & 17.07 & 42.8 & 16.7 & 17.9 & 1.05e+54 & 3.61e+54 & 1.26e+08 \\
    Hole 8 & 0.78 & 19.97 & 55.8 & 9.6 & 12.3 & 7.57e+53 & 3.66e+54 & 1.55e+08 \\
    Hole 9 & 0.56 & 20.51 & 31.2 & 5.7 & 11.5 & 5.31e+53 & 1.13e+54 & 3.69e+07 \\
    \hline
    
    \end{tabular}%
  \label{tab:GF_derived}%
\end{table*}%

\section{Results}
\label{sec:results}
\begin{figure}[!htb]
    \centering
    \includegraphics[width=0.95\linewidth]{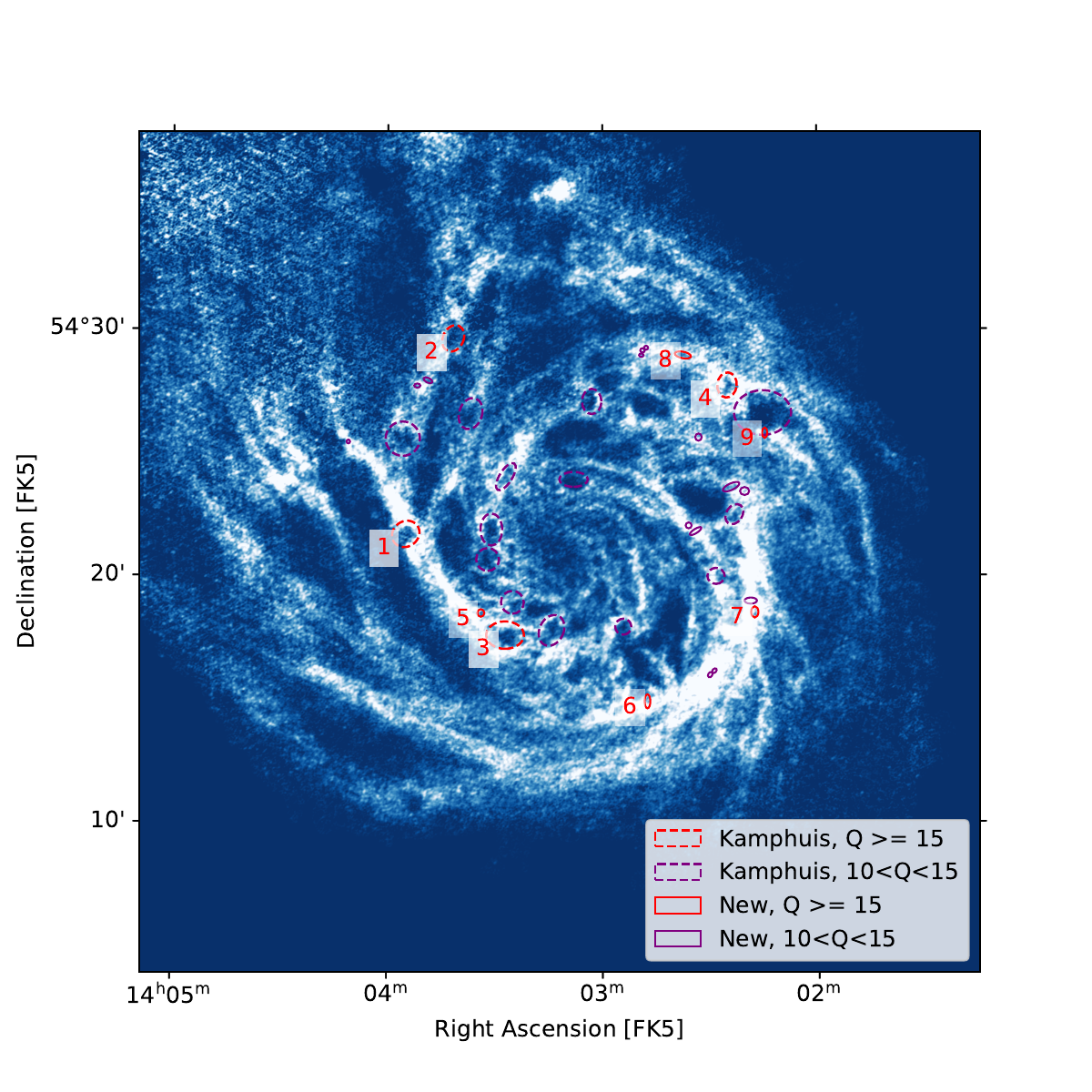}
    \caption{\ion{H}{i} moment-0 map of NGC 5457 (M101), overlaid with the locations of highly likely sites of galactic fountains. Regions are categorized based on quality flags ($Q$): red dashed ellipses mark holes from Kamphuis et al. with $Q \ge 15$, red solid ellipses show newly identified holes with $Q \ge 15$, purple dashed ellipses correspond to Kamphuis et al. holes with $10<Q<15$, and purple solid ellipses mark new detections with $10<Q<15$. The background blue-scale image shows the integrated \ion{H}{i} emission, with brightness scaled between the 10th and 98th percentiles. The nine regions in red are the identified sites of galactic fountains.}
    \label{fig:GF_sites}
\end{figure}
Out of the 72 \ion{H}{i} holes surveyed (52 from \citet{1993Kamphuis} plus 20 newly identified cavities), nine satisfy our quality threshold ($Q\ge15$) and exhibit the most compelling multi-wavelength signatures of galactic fountains. Figure \ref{fig:GF_sites} maps these nine high-confidence sites onto the H I moment-0 image of M101 as red ellipses. Four of the 52 ``classic" Kamphuis holes (dashed, red ellipses) and five of the new bubbles (solid, red ellipses) pass our scoring cut, while holes with intermediate scores ($10<Q<15$) are shown in purple for completeness.

Table \ref{tab:GF_physical} and Table \ref{tab:GF_derived} list the observed and derived properties of these nine holes, respectively. Briefly, their diameters span $0.5-2.4$ kpc (median 0.7 kpc), expansion velocities run from 7 to 54 km s$^{-1}$ (median 10 km s$^{-1}$), and kinetic ages range from 18 to 93 Myr (median $\sim$ 31 Myr). The implied \ion{H}{i} masses displaced are $1 \times 10^7 - 2 \times 10^9 M_\odot$ (median: $1.5\times 10^8 M_\odot$), and the required energies (via Chevalier’s formula) lie between $10^{53}$ and $10^{56}$ erg (median: $3.5\times 10^54$ erg). These best-score holes tend to coincide with prominent spiral-arm \ion{H}{ii} complexes (e.g. NGC 5461, 5462, 5455), confirming that fountain-driving feedback is concentrated in the disk’s active star-forming regions.

\begin{figure*}[!htb]
    \centering
    \includegraphics[width=0.95\linewidth]{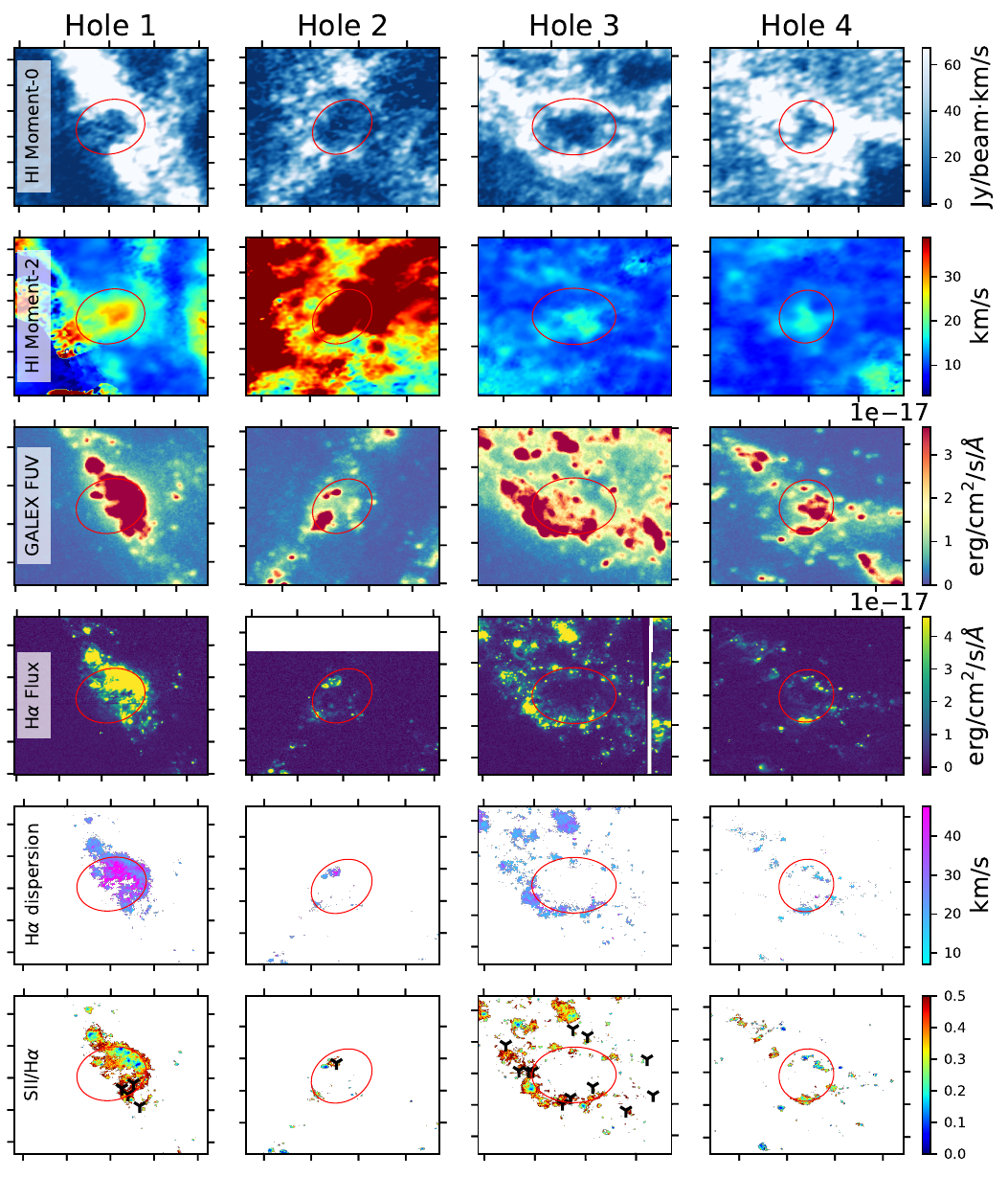}
    \caption{Multiwavelength views of the four biggest sites of galactic fountains in M101 marked by red ellipses. From top: \ion{H}{i} velocity map overlayed with its 5$^\mathrm{th}$ and 95$^\mathrm{th}$ flux contours in black; \ion{H}{i} velocity dispersion; FUV flux map; H$\alpha$ flux map; H$\alpha$ velocity dispersion map; the ratio of [\ion{S}{ii}] to H$\alpha$ flux overlayed with the positions of optically identified SNRs by \citet{1997Matonick} in black points.  }
    \label{fig:big_gf}
\end{figure*}

Our two discovery routes yield a complementary population. The four largest cavities (diameters $1.6-2.4$ kpc; Fig. \ref{fig:big_gf}) were all catalogued by \citep{1993Kamphuis}. In contrast, the five smaller holes ($0.5-0.7$ kpc; Fig. \ref{fig:small_gf}) emerged only from our SITELLE-first approach. This hierarchy of scales demonstrates that high-resolution multi-wavelength data can reveal more modest but still fountain-driven structures that previous \ion{H}{i} surveys missed. For these smaller bubbles, we could not reliably assign Brinks types via PV diagrams; instead, we adopted the local \ion{H}{i} velocity dispersion as an upper limit on the expansion speed.

Hole 3 is the largest cavity (2.4 kpc diameter) and also the oldest ($\sim93$ Myr), located on the south-eastern arm adjacent to NGC 5461. Its PV diagram shows a fully “blown-out” Type I signature, with no coherent shell walls ($v_\mathrm{exp} \lesssim 13$ km s$^{-1}$). The lack of interior H$\alpha$ or FUV emission is consistent with its advanced age (the progenitor OB stars appear to have died). The filamentary clumps at its rim and a two-component fit to the H$\alpha$ line (whose centroid separation matches $v_\mathrm{exp}$), point to triggered star formation along the shell edge. This is also adjacent to where the 2023 SN went off.

Hole 1 is the second largest cavity (2.03~kpc diameter) and the largest expansion velocity of 54 km s$^{-1}$. It was already discovered by \citet{1991Kamphuis} as the first clear case of an expanding \ion{H}{i} shell associated to an \ion{H}{i} hole,
corresponding to the energy of a thousand supernovae.
Situated on the north-eastern arm, Hole 2 exhibits the second largest expansion velocities, $\sim40$ km s$^{-1}$, and was one of the holes inferred from both approaches. Although bubble morphology is evident on the rim of this hole with SITELLE RGB, the H$\alpha$ line does not decompose cleanly into two components, suggesting geometric projection effects or rapid shell fragmentation at this evolutionary stage. Considering the exceptionally high velocity dispersion in this hole (cf second row and second column
in Fig. \ref{fig:big_gf}, and middle panel in Fig. \ref{fig:PVtypes}), it is possible that this hole is suffering in addition from a tidal interaction from
the M101 dwarf companion NGC~5477, as proposed by \citet{1991Combes}, or a collision with a large
extra-galactic gas cloud as proposed by \citet{vanderHulst1988}, since it corresponds exactly to the perturbed region.

Among the small bubbles, Hole 6 stands out as the sole \ion{H}{i} cavity that coincides precisely with an optical rim. With a diameter of just 0.6 kpc, it shows elevated velocity dispersion in both \ion{H}{i} moment-2 and H$\alpha$ $\sigma$ maps at its core, and hosts a compact FUV cluster at its center. 

We note that several new \ion{H}{i} holes rarely coincide with the bubble feature visible in SITELLE RGB imagery. This might be due to the different time scales involved, and also to the previous star formation site and consequent supernovae having triggered new star formation in the adjacent region. Another simpler explanation would be the fact that the part of the disk that pushed away the neutral gas (thereby creating cavities) has now moved due to the rotational motion.

\section{Discussion}
\label{sec:disc}
\subsection{20 New \ion{H}{i} Holes}

\begin{figure*}[!htb]
    \centering
    \includegraphics[width=0.95\linewidth, height=7cm]{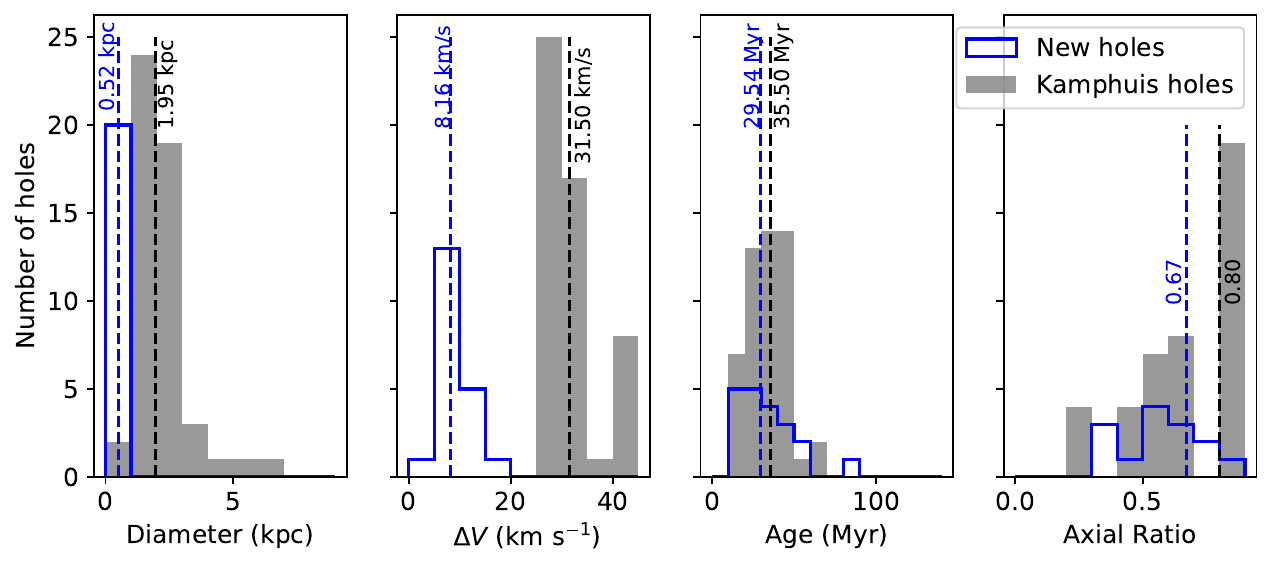}
    \caption{Comparison of properties between the 20 new \ion{H}{i} holes in blue and the 52 Kamphuis holes in grey. The medians of each sample are plotted as dashed lines. As expected, the new holes span smaller diameters and are younger than those of the Kamphuis sample. However, these younger holes do not seem to have a more circular shape than the older ones.}
    \label{fig:hist_diameter}
\end{figure*}
Approach 2 yields 20 new and smaller holes that were missed by the 15'' resolution of the Kamphuis studies. Here, we discuss their size distribution, age, and shape using Fig.~\ref{fig:hist_diameter}. The new holes peak at smaller radii (150-400 pc) whereas the original sample is biased towards radii $\ge 400$ pc. This size‐distribution shift directly reflects the factor‐of 2 improvement in spatial resolution from THINGS (6") versus the earlier 15" WSRT data and demonstrates how higher resolution reveals a hierarchy of feedback structures. 

\begin{figure}[!htb]
    \centering
    \includegraphics[width=0.95\linewidth]{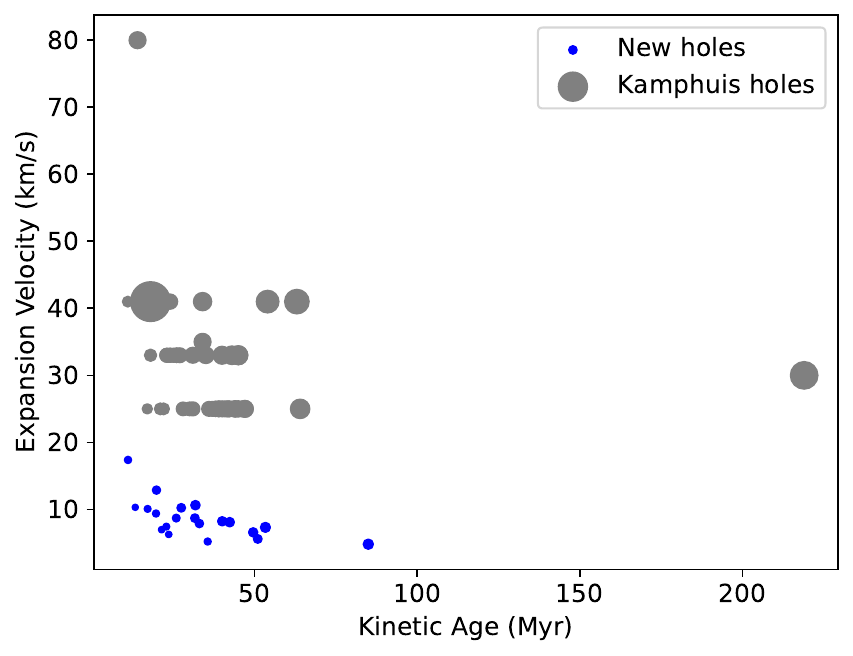}
    \caption{Expansion velocities of the new and old holes plotted against their kinetic age. The size of any symbol is proportional to the diameter of the \ion{H}{i} hole it represents. Even if their sizes are bigger, the relatively higher expansion velocities of the Kamphuis sample holes push them to lie within a younger age area of the plot. The slower expansion velocities, close to the local \ion{H}{i} velocity dispersion ($\sim10$ km s$^{-1}$), render the kinetic age of the new sample 
    very similar to that of the old.}
    \label{fig:age_expansion}
\end{figure}

These new holes expand at only 5-17 km s$^{-1}$, far slower than the 25-45 km$^{-1}$ typical of the Kamphuis sample. Because the PV diagrams of sub-kpc cavities rarely show distinct approaching/receding bumps, we estimate each small hole’s expansion velocity simply as the HI moment-2 dispersion averaged over its area. This procedure naturally yields lower $v_\mathrm{exp}$ than for the Kamphuis holes, due to the measurement method: tiny shells cannot be kinematically resolved as well as 1 kpc-scale bubbles. Consequently, if we compute the $t_\mathrm{kin}$ (as we do for the high Q sites), they appear systematically similar in age (median $\sim30$ Myr) to their larger, faster Kamphuis holes ($\sim 35$ Myr). This is clear from Fig.~\ref{fig:age_expansion} which confirms that we have an observational bias of lower velocities for the newly identified holes.
 We note that the error bars on $v_\mathrm{exp}$ for these small holes are now much larger,
 since the estimated expansion velocity is very close to the \ion{H}{i} typical velocity dispersion of 10~km~s$^{-1}$. These
 large relative errors of up to $\sim$ 100\% prevent us to estimate their corresponding energy with precision.

\begin{figure}[!htb]
    \centering
    \includegraphics[width=0.95\linewidth]{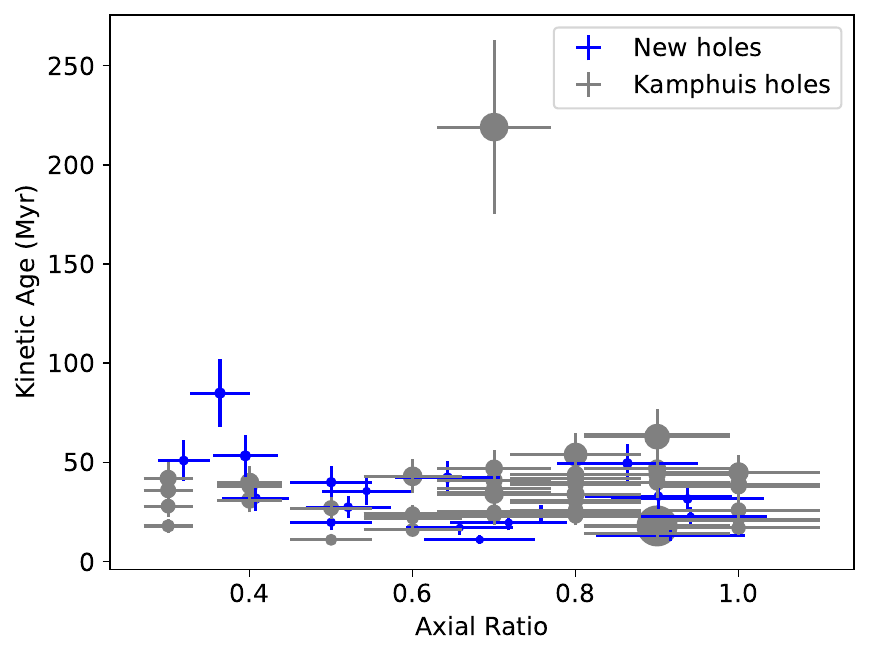}
    \caption{Kinetic age vs axial ratio of the new (blue) and old (grey) holes, with the size of the symbols being representative of the hole diameter. This denotes a very weak negative correlation between axial ratio and kinetic age in our sample. This statistically insignificant relationship confirms that one cannot correlate the shape of the holes to its age.}
    \label{fig:ratio_age}
\end{figure}

Both samples show a wide range of axial ratios (0.4-1.0). Ideally, we expected that the differential rotation will shear holes into an elliptical shape with time. This would also mean that the shape of the bigger holes of the Kamphuis sample should be more elliptical than the younger ones found in this work. Consequently, we compared the axial ratio and kinetic age of holes (Fig.\ref{fig:ratio_age}) and found no statistically significant correlation. This lack of correlation might be due to the low shear in the rotation curve of this late-type galaxy, with a very light bulge. The rotation is
almost a solid-body one until a radius of 10~kpc \citep{Bosma1981}. 

\begin{figure}[!htb]
    \centering
    \includegraphics[width=0.95\linewidth]{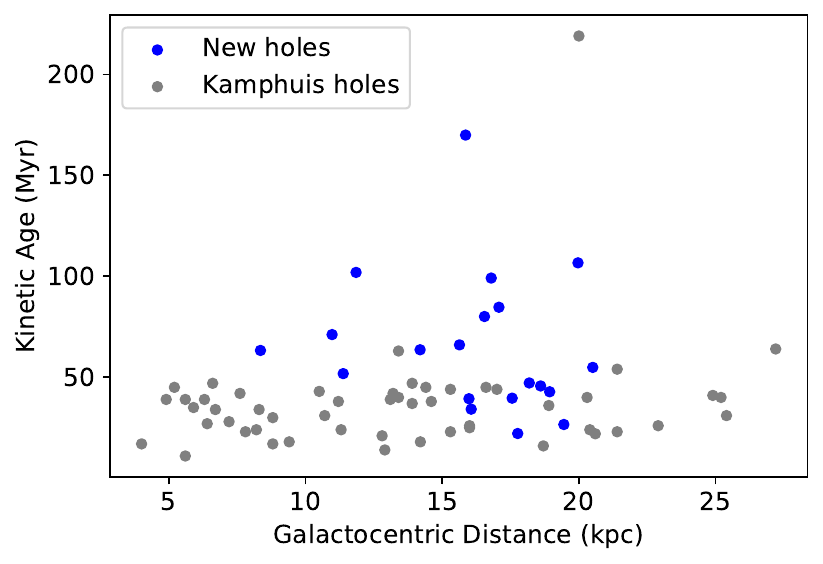}
    \caption{Variation of age with galactocentric distance. The Pearson $r$ value $=0.2$ with a $p-$value of $0.08$. This indicates a weak positive linear correlation between age and galactocentric distance. In other words, as galactocentric distance increases, age tends to increase slightly, but the relationship is not strong.}
    \label{fig:age_gd}
\end{figure}

We also look at the properties of these newly found holes as a function of the radius from the center of the galaxy. There is a weak trend suggesting that the age of the HI hole increases with galactocentric distance. This indicates that inner-disk shells are, on average, younger---consistent with more rapid dissolution in regions of denser gas. However, the scatter is large, and the trend is not yet robust, as shown in Fig.~\ref{fig:age_gd}.

\begin{figure}[!htb]
    \centering
    \includegraphics[width=0.95\linewidth]{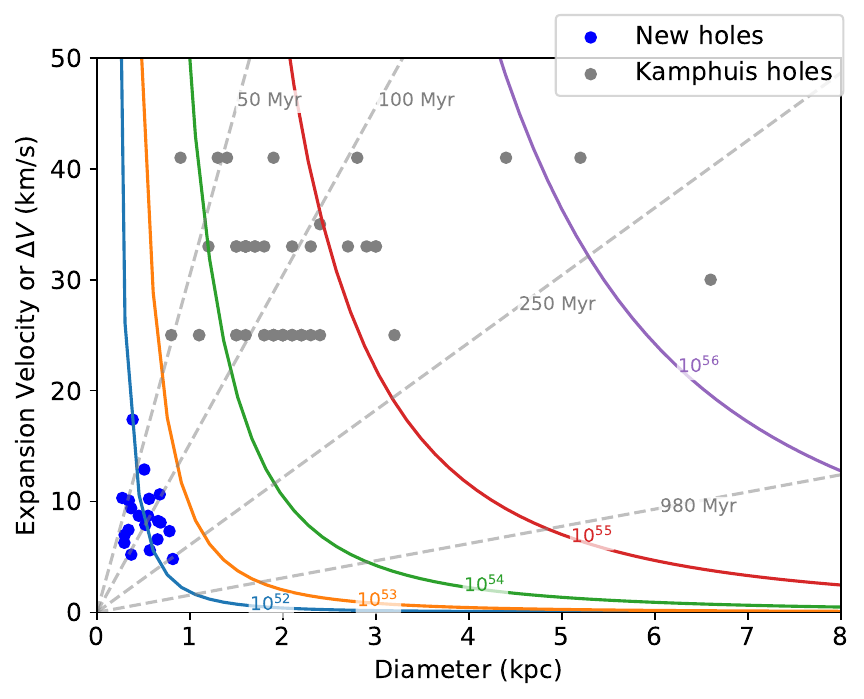}
    \caption{The upper limits of the expansion velocity (given by the average velocity dispersion inside the hole) vs their diameters. The coloured curves are traced assuming the holes are powered by supernovae, with total energy labelled on each curve.
    The dashed grey lines correspond to different ages (see text).}
    \label{fig:vel_d}
\end{figure}

If we assume that all of these holes were indeed powered by Type II supernovae, we can plot their ``expansion'' velocities (which in our case is an upper limit) against their diameters as done in Figure.~\ref{fig:vel_d} and reversely use Eqn.~\eqref{eqn:Ech} to determine the energy required to power them. We take the $n_0$ value to be 0.35 cm$^{-3}$ which is the midplane density discussed by \citet{2003Kuntz}. The following equation is based on the hydrodynamical description of the evolution of supernovae modeled by \citet{1974Chevalier}, where they related the time with the radius (R) of the holes and expansion velocity $v_\mathrm{exp}$ to be: 
\begin{equation}
    t [\text{Myr}] = 0.304R /v_\mathrm{exp}.
\end{equation}
We use this to draw constant time and energy lines on Fig.~\ref{fig:vel_d} as was done by \citep{1992Puche} for the nearby dwarf galaxy Holmberg II. As in their case, this equation gives ages 3 times lower than the kinetic ages we calculate. Furthermore, it is important to note here that these  estimates of energy are not necessarily related to the energy estimated in Table \ref{tab:GF_derived} to power the four largest fountain-sites. 

\begin{figure}[!htb]
    \centering
    \includegraphics[width=0.95\linewidth]{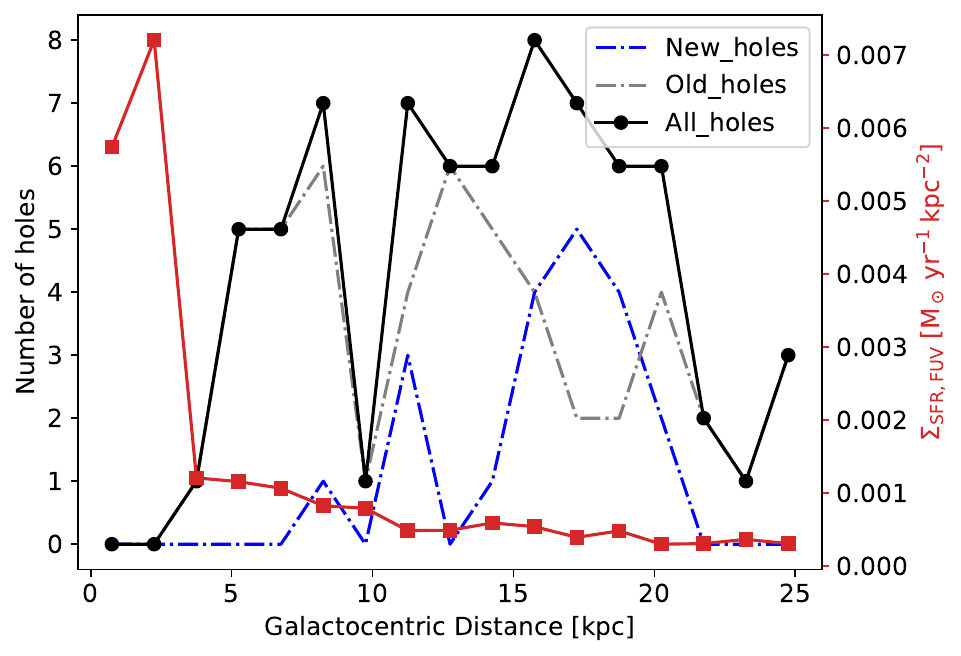}
    \caption{Radial distribution of \ion{H}{i}-hole counts compared to FUV-derived, dust-corrected star-formation-rate surface density. The two profiles exhibit a statistically significant moderate anti-correlation (Pearson $r$ = –0.57, $p$ = 0.017), indicating that regions of elevated recent star formation tend to host fewer HI holes. }
    \label{fig:radial_sfr}
\end{figure}

As a final step, we study the radial trends of the distribution of both the new and old \ion{H}{i} holes in the galaxy as opposed to the star formation rates (SFR). Fig.~\ref{fig:radial_sfr} compares the radial surface density of \ion{H}{i} holes to the dust-corrected FUV SFR profile (Refer to Appendix \ref{app:FUV-SFR} for FUV-derived SFR calculation). We find a moderate anti-correlation ($r=-0.57, p=0.017$) between SFR surface density, $\Sigma_\mathrm{SFR}(R)$ and surface density of the \ion{H}{i} holes. Regions with highest ongoing star formation (inner 3-5 kpc) surprisingly hosts fewer large \ion{H}{i} cavities, perhaps because vigorous feedback has already disrupted old shells or because high SFR goes hand in hand with high gas density, which refills holes more rapidly. Beyond 5 kpc, the number of holes rises as the SFR surface density tails off, suggesting that past episodes of clustered supernovae have carved long-lasting cavities in the outer disk.%as was observed by \citet{2011Bagetakos}

\subsection{Fountains in Face-on Galaxies}
The combination of shell morphology, stellar associations, and kinematic evidence builds a compelling case that stellar feedback is responsible for carving these cavities in M101’s neutral gaseous disk. Comparing this to the edge-on galaxies where one usually detects extraplanar gas or line-of-sight lagging halos as evidence of fountain flows \citep[e.g.][]{Rand1998}, here we see the imprints of those flows on the disk and the launch sites of the fountains. This proves that studying face-on galaxies offers a complementary perspective.  

All nine high-quality fountain candidates coincide with regions of recent star formation. In each case, we detect H$\alpha$ emission within the cavity or along its rim. The former case indicates that the hole is formed in the last few Myrs and the responsible stars are still alive. In the latter case, where H$\alpha$ is rim-dominant, they are almost always coincident with FUV emissions in the centre. From this, we infer that the hole was created hundreds of Myr ago: the ionizing population has aged, yet secondary ``contagious" star formation along the compressed shell continues to power localised H$\alpha$ along the rims. In every fountain site, the supernovae have successfully displaced the neutral gas, revealing clear depressions in \ion{H}{i} moment-zero map, accompanied by elevated velocity dispersion in both \ion{H}{i} and H$\alpha$ (moment-2).

Re-examining Fig.~\ref{fig:big_gf}, the four largest cavities exhibit classical supershell or superbubble morphologies. Two of these (leftmost and rightmost panels) were previously classified as a superbubble and a supershell by \citet{1991Kamphuis} and \citet{2011Chakraborti}, respectively.  Their implied energy requirements $\sim 10^{56-57}$ erg exceed by two orders of magnitude the budget of our smaller fountain sites.

\citet{1997Matonick} carried out an optical survey of M101 and catalogued 93 supernova remnants (SNRs), classifying their morphologies as stellar, filled, diffuse, arc or shell. When we overlay their positions on the [\ion{S}{ii}]/H$\alpha$ map (bottom panel of Figs.~\ref{fig:big_gf} and \ref{fig:small_gf}), five of our nine high-quality fountain candidates lie within a few hundred parsecs of these SNRs. In particular, Holes 1 and 3 are delineated by multiple remnants and additionally, several of our intermediate Q-score detections (purple ellipses) also coincide with SNRs along their rims. 

The multi-wavelength dataset of these nine sites does not show any strong evidence of inflowing gas. However, we can deduce that from other observations of edge-on galaxies, where we see gas above the disk, and also from the amplitude of the expansion velocity, which is far below the escape velocity estimated to be 240 km s$^{-1}$\citep{2003Kuntz}, the gas expelled by the supernovae (at median velocities of 34.5 km s$^{-1}$) must return to the disk. To precise more the maximum height at which the \ion{H}{i} gas can be expelled, and the time-scale of its travel outside the plane, we have computed these quantities, from $v_\mathrm{exp}$ and the restoring force from the stellar disk, in Table \ref{tab:Zmax_Ttot}. This gives an order
of magnitude of the feedback efficiency, when the gas is unavailable for star formation.
\begin{table}
    \centering
    \begin{tabular}{c|c|c}
    \hline
        Hole & Max. Height [pc]  & Time of flight [Myr]  \\
        %&  &  \\
        \hline
 1  &  985.03  &   36.48 \\
 2  & 1767.22  &   86.21 \\
 3  &  134.19  &   20.64 \\
 4  & 1368.89  &   97.78 \\
 5  &   93.14  &   20.70 \\
 6  &  219.22  &   43.84 \\
 7  &  253.94  &   63.49 \\
 8  &  322.58  &   92.16 \\
 9  &  444.55  &   98.79 \\ \hline
    \end{tabular}
    \caption{Maximum vertical height $Z_{max}$ from the $Z_0$ formula \eqref{eqn:z0} by replacing $v_\mathrm{disp}$ with $v_\mathrm{exp}$, and the total time of fountain flight as $2Z_{max}/v_\mathrm{exp}$.}
    \label{tab:Zmax_Ttot}
\end{table}
\subsection{Mechanical Energy}

All of the mechanical energy imparted to the ISM comes from OB stars via ionizing photons, stellar winds or supernovae explosions \citep{1982Abbott, 1987McCray, Agertz2013, Kim2015}.

Of the two energy estimates in this study, we have shown in Table \ref{tab:GF_derived} that both equations from \citet{1974Chevalier} and \citet{1987McCray} give comparable results, in spite of their different sensitivity to expansion velocities and diameters. It is possible to use these estimates to calculate the number of supernovae that produced these features. According to the formula by \citet{1987McCray}, the radius of an \ion{H}{i} shell is given by:

\begin{align}
    R_s \text{[pc]} &= 97 \left( \frac{N_* E_{51}}{n_0}\right)^{1/5} t_7^{3/5}, \\
    N_* &=  \frac{R_s^5 n_0}{97^5t_7^3 E_{51}}
\end{align}
where $N_*$ is the number of supernovae, $t_7$ is the age of the shell in $10^7$ years, $E_{51}$ is the energy deposited by the supernovae in units of $10^{51}$ erg and $n_0$ is the density of the ambient interstellar medium in $cm^{-3}$ which can be considered to be equal to $n_H$ as we did in Section.\ref{sssect:derived}. We found $N_*$ between 100 and $10^4$, for the 9 holes considered
here in detail.

\section{Conclusions}

We have investigated galactic fountains in the almost face-on, spiral galaxy M101, comparing ionized and neutral gas,
mapped through H$\alpha$, UV, \ion{H}{i}-21cm. We use high-resolution data from THINGS for 
the \ion{H}{i} emission, GALEX for UV, and SITELLE/SIGNALS IFU for the H$\alpha$ tracer. The H$\alpha$ gas is ionized by OB stars of life-time shorter than 10 Myr, and the FUV emission remains after 100 Myr, dating the last star forming event.
Supernovae explosions accompanying the star formation blow away the neutral gas, digging \ion{H}{i}-holes in the galaxy plane, and sending material above the plane, which then infalls back to the plane like a fountain \citep{Fraternali2006, Fraternali2008}. This delays star formation, and plays a feedback role. We revisited the 52 \ion{H}{i} holes catalogued by \citet{1993Kamphuis}, and found 20 more holes, essentially in the outer parts of the galaxy. We have characterized the feedback processes and quantified their efficiency, through measurements of sizes, ages, mechanical energies, displaced gas masses. These 20 new holes are smaller in size and expansion velocity, mainly due to the increased spatial and velocity resolution. There is a deficit of small holes in the inner regions. This could be a selection bias, since small holes are easier to detect in the outer parts, where the gas density is lower, while in the inner parts, the small holes are more rapidly refilled.
All \ion{H}{i} hole properties, like their size, axial ratio, 
their atomic gas column density and their kinetic energy, based on the expansion velocity $v_\mathrm{exp}$ of the bubble,
are studied as a function of their distance to the galaxy center. There is no correlation between axis ratio and distance from
the center, may be because this late-type galaxy has almost solid-body rotation until 10~kpc, meaning little
shearing of the bubbles.
Their age is estimated from their size and $v_\mathrm{exp}$, assumed constant during the bubble expansion. 
The energy of the bubble is computed from $v_\mathrm{exp}$, its diameter, and the volumic density 
of the gas $n_0$ in which it expands, from two methods \citep{1974Chevalier, 1987McCray}. We derive from these energies that the holes were created by a number of supernovae between 100 and 10$^4$.
We mapped in more detail the nine holes satisfying strong criteria to be true fountain
effects. Most of holes contain FUV emission inside, and some H$\alpha$ at their borders, or somewhat offset, indicating contagious star formation. Young stars and H$\alpha$ are found inside only one small hole, showing that these are relatively rare. The largest hole of 2.4 kpc and oldest age (94 Myr) has no H$\alpha$ nor UV emission, and is consistent with an old fountain effect, when the star formation has terminated. Globally, the holes are relatively old, and it is difficult to find in their center the 
very star formation at the origin of the hole. Therefore, no correlation can be found between the present star formation and the mechanical energy of the hole.
The kinetic age of \ion{H}{i} holes is remarkably constant with radius, around 50 Myr, and does not vary with size 
and $v_\mathrm{exp}$. The number of holes is smaller in the galaxy center, where star formation is higher, which may look as a paradox, but is in fact a selection bias. It is more difficult to see \ion{H}{i} holes where the gas is denser, and the rotation period is shorter, since holes are then smaller, and destroyed by star formation feedback. 
Finally, we note that some regions might also be perturbed by tidal interactions, as is likely for Hole 2, since M101 is 
highly perturbed, as obvious from its remarkable lopsidedness.

%%%%%%%%%%%%%%%%%%%%%%%%%%%%%%%%%%%%%%%%%%%%%%%%%%%%%%%%%%%%%%
\begin{acknowledgements}
AS would like to thank Laurent Drissen and Thomas Martin for their valuable help with analysing the SITELLE data.
This research was based on observations obtained at the CFHT, which is operated from the summit of Mauna Kea by the National Research Council of Canada, the Institut National des Sciences de l’Univers of the Centre National de la Recherche Scientifique of France, and the University of Hawaii. The authors wish to recognize and acknowledge the very significant cultural role that the summit of Mauna Kea has always had within the indigenous Hawaiian community. The authors are most grateful to have the opportunity to conduct observations from this mountain.
The observations were obtained with SITELLE, a joint project between Université Laval, ABB-Bomem, Université de Montréal, and the CFHT, with funding support from the Canada Foundation for Innovation (CFI), the National Sciences and Engineering Research Council of Canada (NSERC), Fonds de Recherche du Québec – Nature et Technologies (FRQNT), and CFHT.
The collaboration is grateful to the FRQNT, CFHT, the Canada Research Chair program, the National Science foundation NSF – 2109124, Natural Sciences and Engineering Research Council of Canada NSERC – RGPIN-2023-03487, the Swedish Research Council, the Swedish National Space Board, the Royal Society, and the Newton Fund via the award of a Royal Society-Newton Advanced Fellowship (NAF\textbackslash R1\textbackslash180403), FAPESC, CNPq, FAPESP (2014/11156-4), FAPESB (7916/2015), and CONACyT (CB2015-254132).
\end{acknowledgements}

% Please note that we have included the references below in
% order to compile the document, but we ask you to:
\bibliographystyle{bibtex/aa} % style aa.bst
\bibliography{references} % your references Yourfile.bib

%%%%%%%%%%%%%%%%%%%%%%%%%%%%%%%%%%%%%%%%%%%%%%%%%%%%%%%%%%%%%%%
\begin{appendix}
\renewcommand{\thefigure}{A.\arabic{figure}}
\begin{figure*}[!htb]
    \centering
    \includegraphics[width=0.95\linewidth]{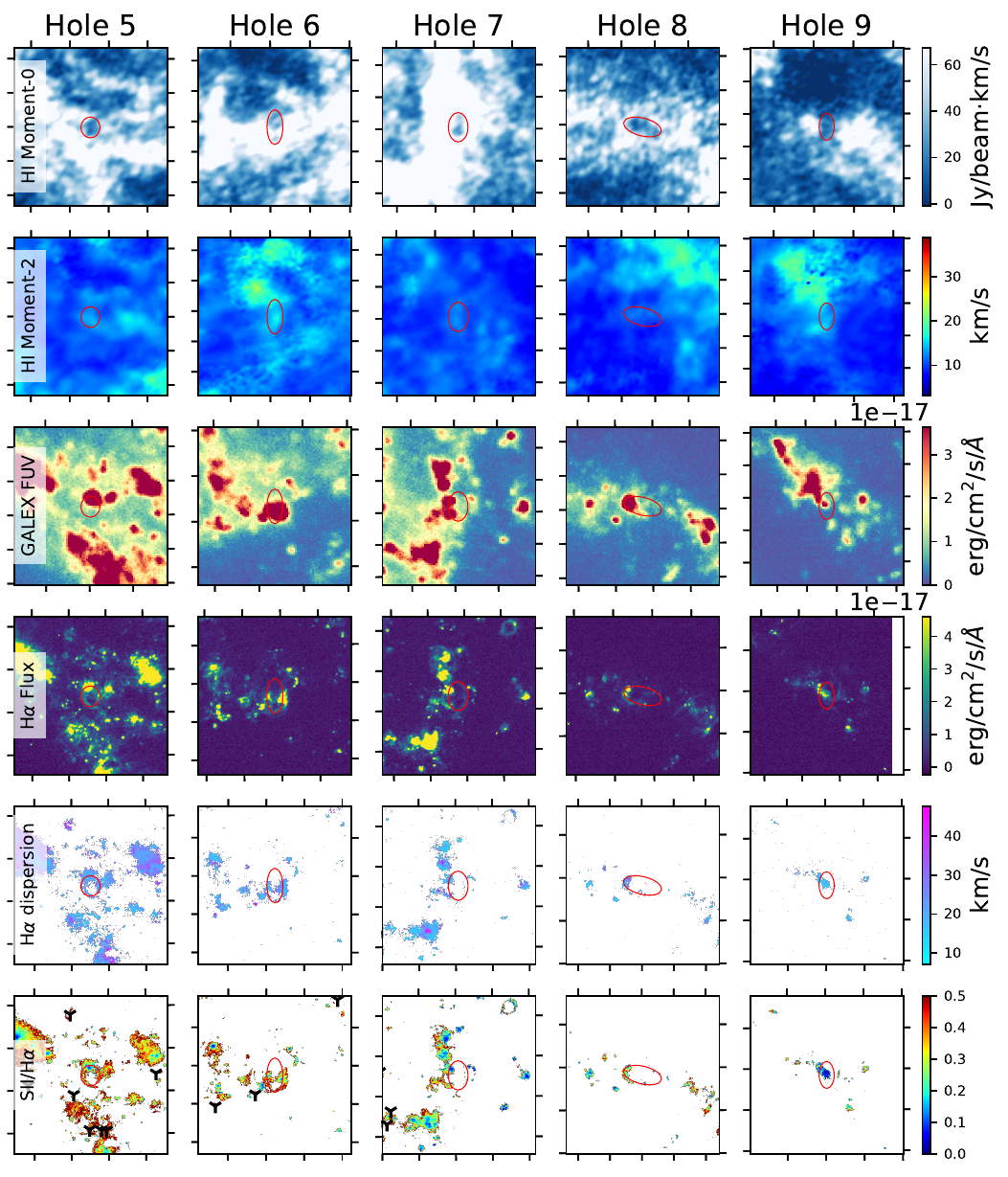}
    \caption{Same as Fig.~\ref{fig:big_gf} for the smaller five sites of galactic fountains in M101.}
    \label{fig:small_gf}
\end{figure*}

\section{SFR calculation using GALEX data}
\label{app:FUV-SFR}

We use the intensity (fd(nd)-int), sky-background (fd(nd)-skybg) and weight (fd(nd)-wt) from the GALEX data products to calculate the radial distribution of SFR of M101. The intensity and the sky-background maps are in units of counts per pixel, while the weight maps are dimensionless. The higher the weight of a pixel, the more confident we are in its intensity value. 

The flux measurements have been made from the background-subtracted versions of the intensity images whose units are in
counts per second (CPS) by multiplying with a factor of $1.40\times10^{15}$ for FUV. This gives the flux,$F_\nu$, in erg/s/cm$^2$/\AA. The FUV luminosity is obtained by:
\begin{equation}
    L_\nu [\mathrm{erg/s}] = 4\pi D_L^2 F_\nu \nu
\end{equation}
where $D_L$ is the luminosity distance to galaxy M101 in cm, $\nu = 442$ is the FUV bandwidth.
This luminosity needs to be corrected for dust attenuation for which we must first calculate the FUV and NUV magnitudes using the background subtracted intensity in counts per second, $I$:
\begin{align}
    m_\mathrm{FUV} &= -2.5 \log_{10} (I_\mathrm{FUV})+18.82\\
    m_\mathrm{NUV} &= -2.5 \log_{10} (I_\mathrm{NUV})+20.08\\
\end{align}

The attenuation in FUV is thus obtained by using the following formula from \citet{2011Hao}:
\begin{equation}
    A_\mathrm{FUV}  = 3.83 \left[ ( m_\mathrm{FUV} - m_\mathrm{NUV}) - 0.022 \right]
\end{equation}
The luminosity is then corrected using 
\begin{align}
    L^\mathrm{corr}_\mathrm{FUV}   =  10^{0.4A_\mathrm{FUV}} L_\mathrm{FUV}
\end{align}
To obtain the dust-corrected SFR from the FUV luminosity, we calculate :
\begin{equation}
    \mathrm{SFR}^\mathrm{corr}_\mathrm{FUV} [M_\odot \mathrm{yr}^{-1}]= \frac{L^\mathrm{corr}_\mathrm{FUV}}  {10^{43.35}}
\end{equation}

% \section{Flux Corrections in SITELLE}
% \label{app:Ha-SFR}

% \begin{table}[!htb]
%     \centering
%     \begin{tabular}{cccc}
%         \hline
%         Field   & SN1   & SN2   & SN3   \\ \hline
%         M101-F1 & 0.59  & 1.001 & 0.919 \\
%         M101-F2 & 0.936 & 0.848 & 0.914 \\
%         M101-F3 & 0.753 & 0.772 & 0.823 \\
%         M101-F4 & 0.761 & 0.809 & 0.907 \\ \hline
%     \end{tabular}
%     \caption{Correction factors applied to SITELLE fluxes in their 3 bands (Sebastien Lavoie, private communication). }
%     \label{tab:flux_corr}
% \end{table}
\end{appendix}

\end{document}